\documentclass[amsmath,amssymb,aps,prd, nofootinbib, superscriptaddress]{revtex4}

\usepackage{graphicx}
\usepackage{dcolumn}
\usepackage{bm}
\usepackage[usenames,dvipsnames]{xcolor}
\usepackage[colorlinks=true,linkcolor=NavyBlue,citecolor=NavyBlue,urlcolor=Blue]{hyperref}
\usepackage[mathlines]{lineno}
\usepackage{amsfonts,color}
\usepackage{tensor}

\newcommand{\tetrad}{\theta}

\newcommand{\lapse}{\alpha}
\newcommand{\shift}{\beta}
\newcommand{\inducedmetric}{\gamma}

\begin{document}

	\title{Primary constraints in general teleparallel quadratic gravity}
	\author{Francesco Bajardi}
	\altaffiliation[]{\href{f.bajardi@ssmeridionale.it}{f.bajardi@ssmeridionale.it}}
	\affiliation{Scuola Superiore Merdionale, Largo S. Marcellino 10, I-80138, Napoli, Italy}
	\affiliation{Istituto Nazionale di Fisica Nucleare (INFN),  sez. di Napoli, Complesso Universitario di Monte S. Angelo, via Cinthia, Ed. N, 80126 Napoli, Italy}
	\author{Daniel Blixt}
	\altaffiliation[]{\href{d.blixt@ssmeridionale.it}{d.blixt@ssmeridionale.it}}
	\affiliation{Scuola Superiore Merdionale, Largo S. Marcellino 10, I-80138, Napoli, Italy}
	\date{\today}

	\begin{abstract}
		The primary constraints for general teleparallel quadratic gravity are presented. They provide a basic classification of teleparallel theories from the perspective of the full nonlinear theory and represent the first step towards a full-fledged Hamiltonian analysis. The results are consistent with the limit of metric and symmetric teleparallel quadratic gravity. In the latter case we also present novel results, since symmetric teleparallel theories have only been partially studied so far. Apart from the general results, we also present the special cases of teleparallel theories classically equivalent to general relativity, which differ by a boundary term from the formulation of Einstein and Hilbert. This affects the constraint algebra as the primary constraints involve a mix of torsion and non-metricity, implying that the symmetries of general relativity are realized in a more intricate way compared to the teleparallel case. In this context, a more detailed understanding will provide insights for energy and entropy in gravity, quantum gravity and numerical relativity of this alternative formulation of general relativity. The primary constraints are presented both in the standard formulation and in irreducible parts of torsion and non-metricity. The special role of axial torsion and its connection to the one-parameter of viable new general relativity is confirmed. Furthermore, we find that one of the irreducible parts of non-metricity affects the primary constraint for shift but not lapse. 
	\end{abstract}
	
		\maketitle
	
		\section{Introduction}
	After more than a hundred years since its conception, General Relativity (GR) has garnered affirmation through various experiments and observations, thus confirming its position as a fundamental piece in comprehending gravity and the Universe. Nonetheless, despite being the most widely accepted theory of gravity, GR exhibits certain deficiencies and limitations \cite{Will:2014kxa}. It clashes with quantum mechanics \cite{Goroff:1985th} and with the related description of gravity at small scales; it fails to wholly elucidate phenomena like dark matter and dark energy, presumed to constitute the bulk of the Universe's composition \cite{Frieman:2008sn}; it predicts the existence of singularities where conventional physics breaks down and lacks a self-consistent theory of quantum gravity aiming to merge GR and quantum mechanics \cite{DeWitt:1967yk, Barack:2018yly}. For all these reasons, throughout the years, the motivations for going beyond GR has grown even stronger \cite{Hehl:1994ue}. Consequently, alternate gravity explanations have emerged, aiming to address specific GR drawbacks such as dark matter and energy, cosmic structure formation, and early-time gravity behavior \cite{Capozziello:2003tk, Nojiri:2010wj, Bamba:2012cp, Sanders:2002pf, Copeland:2006wr, Capozziello:2019cav, Nojiri:2017ncd, Odintsov:2023weg}. 

    Some notable modified gravity theories include \emph{e.g.} modifications to the Einstein-Hilbert gravitational action, by introducing higher-order curvature invariants \cite{Stelle:1976gc, Bajardi:2020mdp, Bajardi:2021hya} or couplings between geometry and scalar fields \cite{Halliwell:1986ja, Uzan:1999ch, Clifton:2011jh, Urban:2020lfk}. The most basic extension, $f(R)$ gravity, incorporates a function of the scalar curvature within the action, resulting in field equations of the fourth order \cite{Sotiriou:2008rp, DeFelice:2010aj, Capozziello:2011et, Starobinsky:2007hu, Nojiri:2006gh, Bajardi:2022ocw}. Some formulations of this theory can yield modifications in the Newtonian potential \cite{Capozziello:2021goa}, thus addressing the Galaxy Rotation Curve issue without relying on Dark Matter, as well as elucidate the Universe's exponential expansion without invoking Dark Energy \cite{Capozziello:2002rd}. In this framework, particular interest was gained by the Starobinsky model \cite{Starobinsky:1980te}, which includes a quadratic term in the scalar curvature to accounts for cosmic inflation.

    Another class GR alternatives breaks the assumption of Levi-Civita connection, which is fundamental in Einstein's gravity in order to obtain metric-compatible connections and fully get the dynamics from the starting line element.  By relaxing the assumption of symmetric connections, for instance, torsion arises in the given space-time \cite{Poplawski:2010kb, Cabral:2019gzh}. In certain instances, this deviation leads to the violation of the Equivalence Principle \cite{Arcos:2004tzt} and offers a means to describe gravity at smaller scales \cite{Casadio:2021zai, Krssak:2015lba}. Particularly, by mandating space-time to be governed solely by torsion rather than curvature, a self-consistent theory of gravity can be developed, mirroring GR's dynamics precisely. This theory, dubbed "Teleparallel Equivalent of General Relativity" (TEGR) \cite{Maluf:2013gaa, Bahamonde:2015zma}, has undergone extensive studies in recent years, becoming the focus of numerous investigations and analysis \cite{Xu:2012jf, Krssak:2018ywd, Obukhov:2002tm, Geng:2011ka}. 

    TEGR represents a theoretical framework that frames gravity as a consequence of torsion in space-time fabric. In this context, the gravitational force is described via a set of tetrad fields (also termed "vierbeins"), forming the basis for space-time geometry depiction. These tetrad fields define a torsion tensor, acting as gravity's source in the theory and representing the antisymmetric part of the Christoffel connection. 

    Within these alternative models, the most studied along with TEGR is the so called "Symmetric Teleparallel Equivalent of General Relativity" (STEGR), based on non-metricity of the space-time. Non-metricity accounts for the possibility that the spacetime might not adhere to the metric compatibility condition, which is a fundamental assumption in GR. As torsion emerges in direct relation to the antisymmetry of the affine connection, non-metricity arises when considering a non-zero covariant derivative of the metric tensor, that is $\nabla_\mu g_{\alpha \beta} \neq 0$. 

    While entirely equivalent to GR in terms of field equations, TEGR and STEGR fails to address the limitations GR poses on larger scales. Hence, akin to $f(R)$ gravity in metric formalism, the Lagrangian density of TEGR can be modified in various ways \cite{Bajardi:2021tul}, such as an arbitrary function of the torsion scalar, giving rise to the so-called $f(T)$ gravity \cite{Cai:2015emx, Li:2010cg}. The latter has been proposed as a solution to late-time issues, like the Universe's accelerated expansion \cite{Ferraro:2006jd, Wu:2010xk}, offering novel solutions and alternative models. However, its capability to better explain gravity's observed behavior than GR remains unclear, necessitating further research to ascertain its viability as a gravitational theory. 

    It is important to note that, despite actions involving $f(R)$, $f(T)$, and $f(Q)$ (with $R$ being the Ricci curvature scalar, $T$ the torsion scalar and $Q$ the non-metricity scalar, as detailed in Sec. \ref{sec:GTQG}) are not in principle equivalent, introducing corresponding boundary terms enables their equivalence \cite{Caruana:2020szx, Bahamonde:2016grb}. 

    The features of $f(T)$ and $f(Q)$ gravity are currently under investigation in literature, particularly in applications within cosmology and astrophysics. For example, in Ref. \cite{Finch:2018gkh} the authors demonstrate that a power-law model of $f(T)$ fits the galaxy rotation curve. On the other hand, Ref. \cite{Aljaf:2022fbk} explores ways to address the $H_0$ tension within $f(T)$ models. In Ref. \cite{Bamba:2010wb} the authors derive an equation of state from $f(T)$ that tackles the dark energy issue. 
    
    In the context of $f(Q)$ gravity, the work in Ref. \cite{Anagnostopoulos:2022gej} studies Big Bang Nucleosynthesis, while Ref. \cite{Capozziello:2022tvv} considers the early stages of the Universe to investigate the slow-roll inflation. In \cite{Bajardi:2020fxh} the authors explore bouncing cosmological models within $f(Q)$ gravity, whereas Ref. \cite{Banerjee:2021mqk} provides wormhole solutions within static and spherically symmetric backgrounds. Also, in Ref. \cite{DAgostino:2022tdk} gravitational waves are studied in the context of $f(Q)$ model, searching for deviations from GR.

    Leading to the same dynamical field equations, GR, TEGR and STEGR are often referred to as "Geometric Trinity of Gravity". However, it is worth pointing out that recent discoveries have shown that there is either strong coupling or ghosts in nonlinear extensions to the trinity of gravity \cite{Gomes:2023tur,Heisenberg:2023wgk}.

    To overcome the latter issue, another possible extension of TEGR, known as "New General Relativity" (NGR), was introduced in \cite{Hayashi:1979qx}. In contrast to $f(T)$, NGR does not involve a nonlinear extension. Instead, it incorporates torsion contractions at the same derivative order as in TEGR. This theory has been restricted to a one-parameter viable theory by requiring the absence of ghosts in its extension of TEGR, based on these assumptions. Additionally, it was discovered that the PPN-parameters align with those of GR, suggesting consistency with solar system tests \cite{Iorio:2012cm}. This theory was found to contain strongly coupled fields \cite{BeltranJimenez:2019nns,Cheng:1988zg} casting doubt if NGR could be considered viable. Recently there has been a renewed interest in NGR \cite{Golovnev:2023jnc} finding that, contrary to previous statements, although not completely problem-free NGR is not generically plagued by ghosts. 

    Based on TEGR, STEGR and their related extensions, it is also possible to consider in the starting gravitational action both torsion and non-metricity, properly contracted as provided in Sec. \ref{sec:GTQG}, giving rise to the so called "General Teleparallel Quadratic Gravity" \cite{BeltranJimenez:2019odq}. Within this geometry it is possible to formulate the ``General Teleparallel equivalent of General Relativity'' (GTEGR). Generally, TEGR, STEGR and GTEGR belong to the so called metric-affine theories of gravity, whose fundamental aspects and applications have been extensively studied in the literature \cite{Hehl:1994ue, Blagojevic:2002du, Blagojevic:1988pp, Blagojevic:2013xpa, Blagojevic:2000qs, Obukhov:2002tm, Obukhov:1996ka, Pasic:2017zwe, Pasic:2005qr, Vassiliev:2001qa, Vassiliev:2003dk}. Despite being probably the least explored sector of metric-affine gravity, GTEGR has shown promise to act as an alternative, or possibly even an improvement, to the notion of energy and entropy in gravity \cite{Gomes:2022vrc}, and this hints towards possible developments in canonical quantization of GR.

    The first approach to face the latter issue is the so called "Arnowitt-Deser-Misner" (ADM) formalism \cite{Arnowitt:1962hi} and emerged as an attempt to address challenges in reconciling GR with Quantum Mechanics. Through a 3+1 decomposition of the metric, the formalism yields a gravitational Hamiltonian and establishes quantization rules, leading to a Schr\"odinger-like equation, known as the Wheeler-De Witt (WDW) equation, initially formulated in \cite{DeWitt:1967yk, DeWitt:1967ub, Wheeler:1957mu}. However, the ADM formalism is no longer seen as the definitive solution for quantizing GR, due to its inability to provide a complete theory of Quantum Gravity. Additionally, it involves an infinite-dimensional superspace that poses challenges for handling. 

    Nevertheless, the 3+1 decomposition represents the very first step towards setting up the Hamiltonian formalism for the given theory and, eventually, find a link between gravity and quantum mechanics. In addition, the restriction of the superspace can be useful in view of applications to quantum cosmology, which in turn has provided interesting insights in describing the early stages of the Universe evolution. Specifically, cosmological restrictions enable a reduction of the configuration superspace to a finite-dimensional minisuperspace, allowing analytical solutions to the WDW equation. Quantum Cosmology, thus, offers insights into the Universe's early stages through the "Wave Function of the Universe", a solution to the WDW equation. Interpreting this wave function is not straightforward due to the absence of a Hilbert space and a definite-positive inner product in gravitational theory, though various interpretations have been proposed \cite{Vilenkin:1988yd, Hawking:1983hj, Vilenkin:1982de, Vilenkin:1984wp, Bousso:2011up}.

    In this paper, we study the 3+1 decomposition and the Hamiltonian constraints of some alternative gravity models, including TEGR, STEGR, GTEGR and their extensions. Specifically the paper is organized as follows: in Sec. \ref{sec:GTQG} we introduce the main properties of TEGR, STEGR, GTEGR and general teleparallel quadratic gravity, to subsequently study the corresponding 3+1 decompositions in Sec. \ref{sec:3+1}. In Secs. \ref{sec:PCtrinity} and \ref{sec:conditionsPC}, the primary constraints of $f(T)$, $f(Q)$ and $f(G)$, namely the extensions of TEGR, STEGR and GTEGR, are considered, as well as the conditions for primary constraints in the irreducible representation of torsion and non-metricity. The shifted algebra among constraints is evaluated in Sec. \ref{sec:Shifted}. Finally, in Sec. \ref{sec:Conclusions} we conclude the work with final considerations and perspectives.

	\section{General teleparallel quadratic gravity}
	\label{sec:GTQG}
     In this section we review the main aspects of those models modifying the assumption of Levi-Civita connection, namely TEGR, STEGR and GTEGR. We also introduce the action for general teleparallel quadratic gravity, which is an extension to GTEGR and the main theory studied in this article. As widely known, within non-flat spacetimes, the arrangement of geodesic paths relies on the nature of the connection $\Gamma^\rho_{\,\,\,\, \mu \nu}$. In GR, adopting a metric-compatible and torsionless connection leads to describing the dynamics only by knowing the form of the metric. However, by relaxing these assumptions, it becomes feasible to define two rank-3 tensors associated with the asymmetric part of $\Gamma^\rho_{\,\,\,\, \mu \nu}$ and the covariant derivative of the metric, that is respectively the "torsion tensor" $T_{\,\,\,\, \mu \nu}^{\alpha}$ and the "non-metricity tensor" $Q_{\rho \mu \nu}$, defined as:
     \begin{equation}
     \label{eq:deftors}
T_{\,\,\,\, \mu \nu}^{\alpha} \equiv 2 \Gamma_{\,\,\,\, [\mu \nu]}^{\alpha}\,, \quad Q_{\rho \mu \nu} \equiv \nabla_{\rho} g_{\mu \nu} \neq 0\,.
\end{equation}
Therefore, the most general Christoffel connection including both contributions (in addition with the Levi-Civita contribution) reads 
\begin{equation}
\Gamma^{\rho}_{\,\,\,\,\mu\nu}=\breve{\Gamma}_{\,\,\,\, \mu \nu}^{\rho}+K^{\rho}_{\,\,\,\,\mu\nu}+L^{\rho}_{\,\,\,\,\mu\nu}\,,
\end{equation}
with $\breve{\Gamma}_{\,\,\,\, \mu \nu}^{\rho}$ being the Levi--Civita connection and
\begin{equation}
K^{\rho}_{\,\,\,\,\mu\nu}\equiv\frac{1}{2}g^{\rho\lambda}\bigl(T_{\mu\lambda\nu}+T_{\nu\lambda\mu}+T_{\lambda\mu\nu}\bigr)%
                                      =-K^{\rho}_{\,\,\,\,\nu\mu} \,,
\end{equation}
\begin{equation}
L^{\rho}_{\,\,\,\,\mu\nu}\equiv\frac{1}{2}g^{\rho\lambda}\bigl(-Q_{\mu \nu \lambda}-Q_{\nu \mu \lambda} + Q_{\lambda\mu\nu}\bigr)=%
                                               L^{\rho}_{\,\,\,\,\nu\mu}\,.
\end{equation}
In GR, both the \emph{Contorsion Tensor} $K^{\rho}_{\,\,\,\,\mu\nu}$ and the \emph{Disformation Tensor}
$L^{\rho}_{\,\,\,\,\mu\nu}$ vanish identically. However, the latter is just a possible choice and, depending on the form of the connection, three main models can be considered, namely\footnote{In TEGR and STEGR $\breve{\Gamma}_{\,\,\,\, \mu \nu}^{\rho}$ is not present, even though it is not strictly zero.}:
\begin{equation}
    \begin{array}{l} \mbox{GR} \rightarrow L^{\rho}{ }_{\mu \nu}=K^{\rho}{ }_{\mu \nu}=0\,, \\ \mbox{TEGR} \rightarrow R^\mu{}_{\nu\rho\sigma}(\Gamma) =L^{\rho}{ }_{\mu \nu}=0\,, \\ \mbox{STEGR} \rightarrow R^\mu{}_{\nu\rho\sigma}(\Gamma)=K^{\rho}{ }_{\mu \nu}=0\,,\end{array}
\end{equation}
where $R^\mu{}_{\nu\rho\sigma}(\Gamma)$ is the Riemann tensor. These three theories are completely equivalent at the level of field equations and this can be shown by evaluating the corresponding actions. Specifically, using the definitions
\begin{eqnarray}
&& S^{p \mu \nu} \equiv K^{\mu \nu p}-g^{p \nu} T_{\,\,\,\,\,\,\, \sigma}^{\sigma \mu}+g^{p \mu} T_{\,\,\,\,\,\,\, \sigma}^{\sigma \nu}\,,
\\
&& Q \equiv - \frac{1}{4} Q_{\alpha \mu \nu} \left[- 2 L^{\alpha \mu \nu}+  g^{\mu \nu} \left(Q^\alpha - \tilde{Q}^\alpha \right) - \frac{1}{2} \left(g^{\alpha \mu} Q^\nu + g^{\alpha \nu} Q^\mu  \right)\right],
\\
&& Q_\mu \equiv Q_{\mu \,\,\,\, \lambda}^{\,\,\,\lambda}\,, \quad \tilde{Q}_{\mu} \equiv Q_{\alpha \mu}^{\, \, \, \, \, \, \alpha}\,, \quad T \equiv T^{p \mu \nu} S_{p \mu \nu}  \,,
\end{eqnarray}
the three actions 
\begin{eqnarray}
&&\mathcal{S}_{GR}\equiv \frac{\kappa}{2}  \int d^4x\,\sqrt{-g}\, R +\mathcal{S}^{(m)}\,,\\
&&\mathcal{S}_{TEGR}\equiv \frac{\kappa}{2} \int d^4x\,\sqrt{-g}\,T+\mathcal{S}^{(m)}\,,\label{TGACT}\\ 
&&\mathcal{S}_{STEGR}\equiv \frac{\kappa}{2} \int d^4x\,\sqrt{-g}\,Q+\mathcal{S}^{(m)}\,, \label{stegraction}
\end{eqnarray}
result equivalent up to a four-divergence. The field equations of TEGR, STEGR and the related extensions can be found \emph{e.g.} in \cite{Cai:2015emx, Capozziello:2011hj, Boehmer:2011gw, Wu:2010xk, Finch:2018gkh, Sotiriou:2009xt, BeltranJimenez:2019esp, BeltranJimenez:2019tme, Dialektopoulos:2019mtr} and will not be further examined here. 
Another interesting extension of GR, which deals with both torsion and non-metricity, is GTEGR. The latter can be obtained as a specific subcase of the general teleparallel quadratic gravity action, \emph{i.e.}

	\begin{align}
	\begin{split}
	S_{||}=\frac{1}{2}M_{pl}^2 \int \mathrm{d}^4 x \sqrt{-g}&\left[a_1 T_{\alpha\mu\nu}T^{\alpha\mu\nu}+a_2T_{\alpha\mu\nu}T^{\nu\mu\alpha}+a_3T_\mu T^\mu \right. \\
	& \left. +b_1Q_{\alpha\mu\nu}T^{\nu\alpha\mu}+b_2Q_\mu T^\mu +b_3 \bar{Q}_\mu T^\mu\right. \\
	& \left. c_1Q_{\alpha\mu\nu}Q^{\alpha\mu\nu}+c_2Q_{\alpha\mu\nu}Q^{\mu\nu\alpha}+c_3Q_\mu Q^\mu+c_4 \bar{Q}_\mu\bar{Q}^\mu+c_5Q_\mu\bar{Q}^\mu \right],
	\end{split}
 \label{GTEGRction}
	\end{align}
 by setting the coefficients to 	
	\begin{align}
	&(a_1,a_2,a_3)=\left(\frac{1}{4},\frac{1}{2},-1 \right), \indent (b_1,b_2,b_3)=(-1,1,-1) \\
	&\mathrm{and} \indent (c_1,c_2,c_3,c_4,c_5)=\left(\frac{1}{4},-\frac{1}{2},-\frac{1}{4},0,\frac{1}{2} \right), 
	\end{align}
where $M_{pl}$ represents the Planck mass and we follow the same convention \footnote{In \cite{BeltranJimenez:2019odq} there is an obvious typo, where the coefficients $c_1$ and $c_2$ are the same (realized by the symmetries of non-metricity). Furthermore, the GTEGR coefficients contain a couple of sign mistakes for the mixed terms. This was, however, corrected in \cite{JimenezCano:2021rlu}.} as Ref. \cite{BeltranJimenez:2019odq}, with $T_\mu=T^\alpha{}_{\mu\alpha}$ and $\bar{Q}_\mu=Q^\alpha{}_{\alpha\mu}$.	
  In order to recover TEGR, one must further require $b_i=c_i=0$, whereas for STEGR one has to impose $a_i=b_i=0$. Both subcases are obviously limits of GTEGR taking non-metricity and torsion to zero, respectively. As better detailed in Sec. \ref{sec:3+1}, the above alternative models can be formulated in terms of tetrads and spin connection. In TEGR, for instance, the involvement of the spin connection does not affect the field equations. Consequently, setting it to zero does not have physical implications and allows the theory to be exclusively expressed through the tetrad field. However, in nonlinear modifications of TEGR, this becomes more complex, as a null spin connection leads to the so called Weitzenb\"ock gauge, influencing the options available for choosing the tetrad. For this reason, in the covariant formulation of TEGR, the inclusion of the spin connection is essential. Yet, there is the possibility to adopt a specific gauge where the spin connection vanishes, without altering the degrees of freedom. A detailed discussion on the gauge fixing for the spin connection can be found \emph{e.g.} in \cite{Blagojevic:2023fys, Blixt:2019mkt, Blixt:2022rpl, Blixt:2023kyr, Golovnev:2021omn}.

	\section{3+1 decomposition and primary constraints}

In adopting the 3+1 formalism, our fundamental variables will be $\alpha,\beta^i,h_{ij},L^\mu{}_\nu$, but we will realize primary constraints from the torsion sector, by using Lorentz indices and to transform the indices we use tetrads.
We denote the tetrad and its inverse respectively as $\theta^A{}_\mu$ and $e_A{}^\mu$, thus the torsion and the non-metricity tensors defined in Sec. \ref{sec:GTQG} can be written as:

	\begin{align}
	T^A{}_{\mu\nu}=2 \partial_{[\mu}\theta^A{}_{\nu]}+\theta^B{}_{[\nu}\Gamma^A{}_{\mu]B}, \indent Q_{\alpha\mu\nu}=\partial_\alpha g_{\mu\nu}-2\Gamma^\beta{}_{\alpha(\mu}g_{\nu)\beta},
	\label{TQtetrad}
	\end{align}
where teleparallelism implies that the connection can be cast in the following way \cite{BeltranJimenez:2022azb} 
\begin{align}
\Gamma^\alpha{}_{\mu\nu}=\left(L^{-1}\right)^\alpha{}_\lambda\partial_\mu L^\lambda{}_\nu .
\end{align}

Here, $L^\mu{}_\nu \ \in \ GL(4,\mathbb{R})$ and has sixteen components. In the limit of symmetric teleparallel geometry $L^\mu{}_\nu=\partial_\nu\xi^\mu$ and a coordinate choice making the connection vanish, known as ``the coincident gauge", can always be made. In this way, the components of $\xi^\mu$ can be thought as St\"uckelberg fields,  manifesting that the coincident gauge is physically equivalent to the covariant formulation \cite{BeltranJimenez:2022azb}. Similarly, in metric teleparallel geometries, transforming the coordinate indices of $L^\mu{}_\nu$ to tangent indices with the use of a tetrad, allows to introduce the so-called spin connection $\omega^A{}_{B\mu}=\Lambda^A{}_C \partial_\mu \Lambda_B{}^C$, which depends on Lorentz matrices $\Lambda^A{}_B$. Torsion is then given by
\begin{align}
    T^A{}_{\mu\nu}=2\partial_{[\mu}\theta^A{}_{\nu]}+2\omega^A{}_{B[\mu}\theta^B{}_{\nu]}
\end{align}  
and one can always perform a Lorentz transformation such that the spin connection vanishes \cite{Blixt:2022rpl}. However, in general teleparallel geometries, the coincident gauge cannot be adopted, as it automatically implies vanishing torsion, as follows from the definition \eqref{eq:deftors}.

 The 3+1 decomposition involves constant-time hypersurfaces $\Sigma_t$ and a normalized normal vector $\xi^\mu$ that complies with the condition $\xi_\mu \xi^\mu=-1$. As outlined and detailed in Ref. \cite{Blixt:2020ekl}, this split exclusively applies to spacetime indices, excluding Lorentz indices. The hypersurfaces $\Sigma_t$ form a manifold denoted by spatial indices $i,j,k,...,$ and feature the induced metric $\inducedmetric_{ij}$. Following this division, the tetrads are characterized by
 \begin{align}
	\theta^A{}_0 =\lapse \xi^A+\shift^i \theta^A{}_i,
	\end{align}
	with $\lapse$ being the lapse function, $\beta^i$ the shift vector and 
	\begin{align}
	\xi^A =-\frac{1}{6}\epsilon^A{}_{BCD}\tetrad^B{}_i \tetrad^C{}_j \tetrad^D{}_k \epsilon^{ijk}
	\end{align}
	\label{sec:3+1}
	
	is normal to hypersurfaces of constant time slices. We use a 3+1 decomposition in the ADM formalism, so that the metric takes the well-known form
	\begin{align}
	g_{\mu\nu}=\begin{bmatrix}
	-\alpha^2 +h_{ij}\beta^i\beta^j & h_{ij}\beta^j \\
	h_{ij}\beta^j & h_{ij}
	\end{bmatrix},
	\end{align} 
	where $h_{ij}=\theta^A{}_i\theta^B{}_j\eta_{AB}$ is the induced metric, with $\eta_{AB}$ being the Minkowski metric. For brevity, we present results with noncanonical index positions, like $\beta_i \equiv h_{ij}\beta^j$. However, note that this is a shorthand notation, especially when one for example applies derivatives or variations to such objects. 

	Our canonical variables are $\alpha, \beta^i, h_{ij}, L^\mu{}_\nu$ and to realize the primary constraints we will perform the irreducilbe decomposition of velocities (and their conjugate momenta) under the rotation group $\mathcal{O}(3)$, as has been done in \cite{Blagojevic:2000qs,Blagojevic:2000xd,Blixt:2018znp,Blixt:2019mkt,Blixt:2020ekl,Pati:2022nwi}: 
	\begin{align}
	\dot{L}^A{}_i={}^{\mathcal{V}}\dot{L}_i+{}^{\mathcal{A}}\dot{L}_{ji}h^{kj}\theta^A{}_k+{}^{\mathrm{S}}\dot{L}_{ji}h^{jk}\theta^A{}_k.
	\end{align}
	We here omit the full decomposition and combine the trace with the trace-free part, since this sector is of less interest for primary constraints, as we shall see later in this section.
	
	By making a ADM decomposition of torsion and non-metricity and applying the chain rule, one gets
	\begin{align}
	\overset{\alpha}{\pi}:=\frac{\partial \mathcal{L}}{\partial \dot{\alpha}}=-2\alpha \frac{\partial \mathcal{L}}{\partial Q_{000}},
	\end{align}
	
	\begin{align}
	\overset{\beta}{\pi}{}_i:=\frac{\partial \mathcal{L}}{\partial \dot{\beta^i}}=\frac{\partial \mathcal{L}}{\partial Q_{00}{}^i}+2\beta_i\frac{\partial \mathcal{L}}{\partial Q_{000}},
 \label{shiftconj}
	\end{align}
	
\begin{align}
	\pi^{ij} :=\frac{\partial \mathcal{L}}{\partial \dot{h}_{ij}}= \frac{\partial \mathcal{L}}{\partial Q_{0ij}}+\beta^i\beta^j\frac{\partial \mathcal{L}}{\partial Q_{000}}+\beta^i\frac{\partial \mathcal{L}}{\partial Q_{00j}}+\beta^j\frac{\partial \mathcal{L}}{\partial Q_{00i}}.
	\end{align}

 Next we need to find the conjugate momenta with respect to $L$, namely the rank-2 tensor $P_\mu{}^\nu:=\dfrac{\partial \mathcal{L}}{\partial \dot{L}^\mu{}_\nu}$, which turns out to be 
 \begin{align}
 		\begin{split}
 	P_\mu{}^0&=\overset{\alpha}{\pi}\left(-\alpha\left(L^{-1} \right)^0{}_\mu +\left(L^{-1} \right)^i{}_\mu \beta_{i}-\left(L^{-1}\right)^0{}_\mu\frac{\beta^2}{\alpha}-2\left(L^{-1}\right)^i{}_\mu\frac{\beta_i}{\alpha} \right)\\
 	&-2\overset{\beta}{\pi}_i\left(\beta^i\left(L^{-1}\right)^0{}_\mu+\left(L^{-1}\right)^i{}_\mu\right)=:\overset{\alpha}{\pi}\overset{\alpha}{S}_\mu+\overset{\beta}{\pi}{}_i\overset{\beta}{S}{}^i_\mu.
 	\end{split}
 	\end{align}
 Thus, we get four primary constraints for the connection
\begin{align}
\label{eq:TempLConstraint}
	{}^{P}C_\mu{}^0=P_\mu{}^0-\overset{\alpha}{\pi}\overset{\alpha}{S}_\mu+\overset{\beta}{\pi}{}_i\overset{\beta}{S}{}^i_\mu\approx 0.
\end{align}
The other twelve components of $P$ can be written as
	\begin{align}
\begin{split}
P_\mu{}^i&=\frac{1}{2}\theta^B{}_\mu\left(L^{-1} \right)^A{}_B \frac{\partial \mathcal{L}}{\partial T^A{}_{0i}}-2\pi^{ij}\left(\beta_j\left(L^{-1}\right)^0{}_\mu+\left(L^{-1}\right)^k{}_\mu h_{jk} \right)\\
&+2\overset{\beta}{\pi}{}^{i}\left((\alpha^2-\beta^2)\left(L^{-1} \right)^0{}_\mu-\left(L^{-1} \right)^j{}_\mu \beta_{j}\right)+2\beta^i\overset{\beta}{\pi}_j\left(\beta^j\left(L^{-1}\right)^0{}_\mu+\left(L^{-1}\right)^j{}_\mu\right)\\
&+4\overset{\alpha}{\pi}\left(2\alpha\beta^i\left(L^{-1}\right)^0{}_\mu-\left(L^{-1} \right)^0{}_\mu \frac{\beta^i}{\alpha}\beta^2-\left(L^{-1} \right)^j{}_\mu \frac{\beta^i}{\alpha}\beta_j\right)\\
&=:\frac{1}{2}\theta^B{}_\mu\left(L^{-1} \right)^A{}_B \frac{\partial \mathcal{L}}{\partial T^A{}_{0i}}+\pi^{ij}S_{\mu j}+\overset{\beta}{\pi}{}^i\overset{\beta}{S}_\mu-\beta^i\overset{\beta}{\pi}{}_j\overset{\beta}{S}{}_\mu^j+\overset{\alpha}{\pi}\overset{\alpha}{S}{}_\mu^i.
\end{split}
\end{align}
The vector momenta is obtained by the following particular contraction
\begin{align}
\label{vectconj}
	{}^\mathcal{V}P^i=2\xi^DL^C{}_De_C{}^\mu P_\mu{}^i=\xi^A \frac{\partial \mathcal{L}}{\partial T^A{}_{0i}}
\end{align}
and similarly one obtains the antisymmetric and symmetric momenta
\begin{align}
{}^\mathcal{A}P^{[ji]}=2L^C{}_De_C{}^\mu \theta^D{}_kh^{k[j}P_\mu{}^{i]}=\theta^A{}_kh^{k[j} \frac{\partial \mathcal{L}}{\partial T^A{}_{|0|i]}},
\end{align}

\begin{align}
\begin{split}
{}^\mathcal{S}P^{(ji)}=&L^C{}_De_C{}^\mu \theta^D{}_kh^{k(j}P_\mu{}^{i)}=\frac{1}{2}\theta^A{}_kh^{k(j} \frac{\partial \mathcal{L}}{\partial T^A{}_{|0|i)}}+L^C{}_De_C{}^\mu \theta^D{}_kh^{k(j}\overset{\beta}{\pi}{}^{i)}\overset{\beta}{S}_\mu-L^C{}_De_C{}^\mu \theta^D{}_kh^{k(j}\beta^{i)}\overset{\beta}{\pi}{}_j\overset{\beta}{S}{}_\mu^j\\
&+\overset{\alpha}{\pi}L^C{}_De_C{}^\mu \theta^D{}_kh^{k(j}\overset{\alpha}{S}{}_\mu^{i)}+L^C{}_De_C{}^\mu \theta^D{}_kh^{k(j}\pi^{i)k}S_{\mu k}\\
&=\frac{1}{2}\theta^A{}_kh^{k(j} \frac{\partial \mathcal{L}}{\partial T^A{}_{|0|i)}}+L^C{}_De_C{}^\mu \theta^D{}_kh^{k(j}\overset{\beta}{\pi}{}^{i)}\overset{\beta}{S}_\mu-L^C{}_De_C{}^\mu \theta^D{}_kh^{k(j}\beta^{i)}\overset{\beta}{\pi}{}_j\overset{\beta}{S}{}_\mu^j\\
&+\overset{\alpha}{\pi}L^C{}_De_C{}^\mu \theta^D{}_kh^{k(j}\overset{\alpha}{S}{}_\mu^{i)}+L^C{}_De_C{}^\mu \theta^D{}_kh^{k(j}\pi^{i)k}S_{\mu k}\\
&=:\frac{1}{2}\theta^A{}_kh^{k(j} \frac{\partial \mathcal{L}}{\partial T^A{}_{|0|i)}}+{}^S\Pi^{ij}.
\end{split}
\end{align}
forming the whole set of primary constraints for GTEGR.
 The primary constraints can be obtained from the above conjugate momenta, specifically looking at conditions for which they become independent of velocities. By defining $\overset{\alpha}{\mathcal{A}}_1=c_1+c_2+c_3+c_4+c_5$, $\overset{\alpha}{\mathcal{A}}_2=2c_3+c_5$ and $\overset{\alpha}{\mathcal{A}}_3=b_2+b_3$, the conjugate momentum with respect to the lapse function can be recast as:
	\begin{align}
	\begin{split}
	\overset{\alpha}{\pi}=\sqrt{h}M_{pl}&\left[2\overset{\alpha}{\mathcal{A}}_1\frac{Q_{000}}{\alpha^4}-4\overset{\alpha}{\mathcal{A}}\frac{Q_{00i}\beta^i}{\alpha^4}-2\overset{\alpha}{\mathcal{A}}_1\frac{Q_{i00}\beta^i}{\alpha^4}+4\overset{\alpha}{\mathcal{A}}_1\frac{Q_{i0j}\beta^i\beta^j}{\alpha^4}+2\overset{\alpha}{\mathcal{A}}_1\frac{Q_{0ij}\beta^i\beta^j}{\alpha^4}\right. \\ 
	& \left. -\overset{\alpha}{\mathcal{A}}_2\frac{Q_0}{\alpha^2}-(2c_4+c_5)\frac{\bar{Q}_0-\bar{Q}_i\beta^i}{\alpha^2}-2\overset{\alpha}{\mathcal{A}}_1\frac{Q_{ijk}\beta^i\beta^j\beta^k}{\alpha^4}+\overset{\alpha}{\mathcal{A}}_2\frac{Q_{i}\beta^i}{\alpha^2}\right. \\
	& \left.-\overset{\alpha}{\mathcal{A}}_3\frac{T^A{}_{0i}\theta_A{}^i}{\alpha^2} -\overset{\alpha}{\mathcal{A}}_3\frac{T^A{}_{ij}\beta^i\theta_A{}^j}{\alpha^2} \right].
	\end{split}
	\end{align}
	
	When the torsion vanishes, the above expression is consistent with Ref. \cite{Dambrosio:2020wbi}; moreover, a primary constraint arises iff the R.H.S. is independent of velocities, \emph{i.e.} when the non-metricity term $Q_{0**}$ and the torsion term $T_0$ cancel each other out. As a first approximation, the primary constraint can be simply obtained by setting $\overset{\alpha}{\mathcal{A}}_1=\overset{\alpha}{\mathcal{A}}_2=\overset{\alpha}{\mathcal{A}}_3=0$ resulting in the following constraint: 
	\begin{align}
	\begin{split}
	\overset{\alpha}{C}=\overset{\alpha}{\pi}+\frac{M_{pl}\sqrt{h}}{\alpha^2}\left[\left(2c_4+c_5 \right)\left(\bar{Q}_0-\bar{Q}_i\beta^i \right)\right] \approx 0.
	\end{split}
	\end{align}

	Before writing the explicit expression of the shift conjugate momentum, it useful to define $\overset{\beta}{\mathcal{A}}_1=2c_1+c_2+c_4$ and $\overset{\beta}{\mathcal{A}}_2=b_1-b_3$,  so that Eq. \eqref{shiftconj} can be written as:

	\begin{align}
	\begin{split}
	\overset{\beta}{\pi}{}_i=\frac{M_{pl}\sqrt{h}}{2}&\left[2\overset{\beta}{\mathcal{A}}_1\frac{Q_{00i}}{\alpha^3} -2\overset{\beta}{\mathcal{A}}_1\frac{Q_{0ij}\beta^j}{\alpha^3}-2\overset{\beta}{\mathcal{A}}_1\frac{Q_{j0i}\beta^j}{\alpha^3}-2(2c_2+c_5)\frac{Q_{i0j}\beta^j}{\alpha^3}-c_5\frac{Q_i}{\alpha} \right. \\
	& \left. +(2c_2+c_5)\frac{Q_{i00}}{\alpha^3}+(2c_2+c_5)\frac{Q_{ijk}\beta^j\beta^k}{\alpha^3}+2\overset{\beta}{\mathcal{A}}_1\frac{Q_{jik}\beta^j\beta^k}{\alpha^3}-2c_4\frac{\bar{Q}_i}{\alpha}\right. \\
	& \left. +\overset{\beta}{\mathcal{A}}_2\frac{T^A{}_{0i}\xi_A}{\alpha^2}+\overset{\beta}{\mathcal{A}}_2\frac{T^A{}_{ij}\xi_A \beta^j}{\alpha^2}-b_3\frac{T^A{}_{ij}\theta_A{}^j}{\alpha} \right].
	\end{split}
	\end{align}	
	
	From the above equation, it turns out that the associated primary constraints appear iff $\overset{\beta}{\mathcal{A}}_1=\overset{\beta}{\mathcal{A}}_2=0$, giving
	\begin{align}
	\begin{split}
	\overset{\beta}{C}{}^i=\overset{\beta}{\pi}{}^i-\frac{M_{pl}\sqrt{h}}{2}&\left[\frac{(2c_2+c_5)}{\alpha^3}\left(Q^i{}_{00}+Q^i{}_{jk}\beta^j\beta^k-2Q^i{}_{0j}\beta^j \right)-c_5\frac{Q^i}{\alpha}-2c_4\frac{\bar{Q}^i}{\alpha} -b_3\frac{T^i}{\alpha}  \right]\approx 0.
 \label{Cbeta}
	\end{split}
	\end{align}
	Also here, Eq. \eqref{Cbeta} is consistent with Ref. \cite{Dambrosio:2020wbi}, but it is worth noticing that torsion arises due to the presence of mixed terms. For this reason, it is expected that diffeomorphism will be realized from the Hamiltonian analysis in a more complex way than GR and STEGR. 
	
	Let us now consider the vector irreducible part of the conjugate momentum \eqref{vectconj}, which, by means of the definitions $\overset{\mathcal{V}}{\mathcal{A}}_1=2a_1+a_2+a_3$ and $\overset{\mathcal{V}}{\mathcal{A}}_2=\overset{\beta}{\mathcal{A}}_2=b_1-b_3$, can be recast in the following compact form:
	\begin{align}
	\begin{split}
	{}^{\mathcal{V}}P^i=-\frac{M_{pl}\sqrt{h}}{\alpha^2}&\left[\overset{\mathcal{V}}{\mathcal{A}}_1\alpha\xi_AT^A{}_{0}{}^i-\overset{\mathcal{V}}{\mathcal{A}}_1\alpha\xi_A T^A{}_{j}{}^i\beta^j-a_3\alpha\theta^B{}_j\eta_{AB} T^{Aji}+\frac{\overset{\mathcal{V}}{\mathcal{A}}_2}{2}Q_{00}{}^i\right. \\
	& \left. -\frac{\overset{\mathcal{V}}{\mathcal{A}}_2}{2}\beta^jQ_{0j}{}^i+\frac{b_2}{2}\alpha^2 Q^i+\frac{b_3}{2}\alpha^2\bar{Q}^i-\frac{1}{2}(b_1+b_2)Q^i{}_{00}+(b_1+b_2)Q^i{}_{0j}\beta^j\right. \\
	& \left.-\frac{\overset{\mathcal{V}}{\mathcal{A}}_2}{2}Q_{j0}{}^i\beta^j+\frac{\overset{\mathcal{V}}{\mathcal{A}}_2}{2}Q_{jk}{}^i\beta^j\beta^k-\frac{1}{2}(b_1+b_2)Q^i{}_{jk}\beta^j\beta^k \right].
	\end{split}
	\end{align}	
	By setting $\overset{\mathcal{V}}{\mathcal{A}}_1=\overset{\mathcal{V}}{\mathcal{A}}_2=0$, we obtain the following primary constraint
	
	\begin{align}
	\begin{split}
	\label{eq:VPC}
	{}^{\mathcal{V}}C^i={}^{\mathcal{V}}P^i+\frac{M_{pl}\sqrt{h}}{\alpha^2}&\left[-a_3\alpha\theta^B{}_j\eta_{AB} T^{Aji} +\frac{b_2}{2}\alpha^2 Q^i+\frac{b_3}{2}\alpha^2\bar{Q}^i-\frac{1}{2}(b_1+b_2)Q^i{}_{00}\right. \\
	& \left.+(b_1+b_2)Q^i{}_{0j}\beta^j-\frac{1}{2}(b_1+b_2)Q^i{}_{jk}\beta^j\beta^k \right]\approx 0.
	\end{split}
	\end{align}	
	With regard to the antisymmetric irreducible part of the conjugate momentum, it is enough to define $\overset{\mathcal{A}}{\mathcal{A}}_1=2a_1-a_2$ to get
	\begin{align}
	\begin{split}
	{}^{\mathcal{A}}P^{[ji]}=\frac{\sqrt{h}M_{pl}}{4\alpha}&\left[2\overset{\mathcal{A}}{\mathcal{A}}_1T^A{}_{0}{}^j\theta_A{}^i-2\overset{\mathcal{A}}{\mathcal{A}}_1T^A{}_{0}{}^i\theta_A{}^j+2\overset{\mathcal{A}}{\mathcal{A}}_1T^{Ajk}\beta_k\theta_A{}^i +2\overset{\mathcal{A}}{\mathcal{A}}_1T^{Aki}\beta_k\theta_A{}^j\right. \\
	& \left. +4a_2\alpha \xi_A T^{Aij}-b_1\alpha^3Q^j{}_0{}^i+b_1\alpha^3Q^i{}_0{}^j+b_1\alpha Q^{jik}\beta_k-b_1\alpha Q^{ijk}\beta_k\right]
	\end{split}
	\end{align}
	and the primary constraint occurs as soon as $\overset{\mathcal{A}}{\mathcal{A}}_1=0$, where we have
	\begin{align}
	\begin{split}
	\label{eq:APC}
	{}^{\mathcal{A}}C^{[ji]}={}^{\mathcal{A}}P^{[ji]}-\frac{\sqrt{h}M_{pl}}{4}&\left[4a_2 \xi_A T^{Aij}-b_1\alpha^2Q^j{}_0{}^i+b_1\alpha^2Q^i{}_0{}^j+b_1 Q^{jik}\beta_k-b_1 Q^{ijk}\beta_k\right]\approx 0.
	\end{split}
	\end{align}
		The presence of non-metricity in Eq. \eqref{eq:APC} demonstrates the influence stemming from mixed terms.
Before looking at the symmetric momenta ${}^{\mathcal{S}}P$ it is useful to define the following: ${}^S\mathcal{A}_1=b_1+2a_1+a_2$, ${}^S\mathcal{A}_2=b_2+a_3$, ${}^S\mathcal{A}_3=b_2+b_3$, ${}^S\mathcal{A}_4=b_2+4c_3$, ${}^S\mathcal{A}_5=2c_3+c_5$, ${}^S\mathcal{A}_6=b_1+4c_1$, ${}^S\mathcal{A}_7=c_1+c_2+c_3+c_4+c_5$, and ${}^S\mathcal{A}_8=2c_1+c_2+c_4$. The momenta takes the following form: 
\begin{align}
\begin{split}
	{}^{\mathcal{S}}P^{ij}
	=\frac{M_{pl}}{2\alpha^2}&\left[-2{}^S\mathcal{A}_1\alpha \theta_A{}^jT^A{}_0{}^i-2{}^S\mathcal{A}_1\alpha\theta_A{}^iT^A{}_0{}^j-4{}^S\mathcal{A}_2\alpha T_0h^{ij}  +4{}^S\mathcal{A}_3T_0\frac{\beta^i\beta^j}{\alpha}-2{}^S\mathcal{A}_4Q_0\frac{h^{ij}}{\alpha}\right. \\
	& \left. +4{}^S\mathcal{A}_5Q_0\frac{\beta^i\beta^j}{\alpha}+2({}^S\mathcal{A}_3+2{}^S\mathcal{A}_5)Q_{000}\frac{h^{ij}}{\alpha} -8{}^S\mathcal{A}_7Q_{000}\frac{\beta^i\beta^j}{\alpha}-4({}^S\mathcal{A}_3+2{}^S\mathcal{A}_5)Q_{00k}\frac{\beta^k}{\alpha}h^{ij} \right. \\
	& \left. +16{}^S\mathcal{A}_7Q_{00k}\frac{\beta^i\beta^j\beta^k}{\alpha^3}-4{}^S\mathcal{A}_8Q_{00}{}^i\frac{\beta^j}{\alpha} -4{}^S\mathcal{A}_8Q_{00}{}^j\frac{\beta^i}{\alpha}+2({}^S\mathcal{A}_3+2{}^S\mathcal{A}_5)Q_{0kl}\frac{\beta^k\beta^l}{\alpha}h^{ij}\right. \\
	& \left. -8{}^S\mathcal{A}_7Q_{0kl}\frac{\beta^i\beta^j\beta^k\beta^l}{\alpha^3}+4{}^S\mathcal{A}_8Q_{0k}{}^i\frac{\beta^j\beta^k}{\alpha} +4{}^S\mathcal{A}_8Q_{0k}{}^j\frac{\beta^i\beta^k}{\alpha}-2{}^S\mathcal{A}_6\alpha Q_0{}^{ij}  \right]+{}^SS^{ij}+{}^S\Pi^{ij},
	\end{split}
\end{align}
where the purely spatial part is given by
\begin{align}
	\begin{split}
	{}^{S}S^{ij}=\frac{M_{pl}\sqrt{h}}{\alpha^2}&\left[{}^S\mathcal{A}_1\alpha T^{Akj}\beta_k\theta_A{}^i+{}^S\mathcal{A}_1\alpha T^{Aki}\theta_A{}^j-b_3\alpha T^A{}_{k}{}^i\theta_A{}^k\beta^j  -b_3\alpha T^A{}_k{}^j\theta_A{}^k\beta^i \right. \\
	& \left.  +\alpha h^{ij}T^A{}_{kl}\theta_A{}^l\beta^k\left(2a_3+2b_2\right)-2{}^S\mathcal{A}_3\frac{\beta^i\beta^j\beta^k}{\alpha}T^A{}_{kl}\theta_A{}^l+{}^S\mathcal{A}_4\alpha Q_k\beta^kh^{ij}+\alpha \bar{Q}_k\beta^kh^{ij}\left(b_3+2c_5 \right) \right. \\
	& \left. -({}^S\mathcal{A}_3+2{}^S\mathcal{A}_5)h^{ij}Q_{klm}\frac{\beta^k\beta^l\beta^m}{\alpha}+{}^S\mathcal{A}_6\alpha Q_{k}{}^{ij}\beta^k+\frac{\alpha}{2}Q^{jik}\beta_k\left(b_1+c_2 \right)+\frac{\alpha}{2}Q^{ijk}\beta_k\left(b_1+c_2 \right) \right. \\
	& \left. -2{}^S\mathcal{A}_8Q_{kl}{}^j\frac{\beta^i\beta^k\beta^l}{\alpha}-2{}^S\mathcal{A}_8Q_{kl}{}^i\frac{\beta^j\beta^k\beta^l}{\alpha}+4{}^S\mathcal{A}_7Q_{klm}\frac{\beta^i\beta^j\beta^k\beta^l\beta^m}{\alpha^3}  +Q^j{}_{kl}\frac{\beta^i\beta^k\beta^l}{\alpha}\left(-2c_2-c_5 \right)\right. \\
	& \left. +Q^i{}_{kl}\frac{\beta^j\beta^k\beta^l}{\alpha}\left(-2c_2-c_5 \right)-2{}^S\mathcal{A}_5Q_k\frac{\beta^i\beta^j\beta^k}{\alpha}+2c_4\alpha\bar{Q}^j\beta^i +2c_4\alpha\bar{Q}^i\beta^j +\bar{Q}_k\frac{\beta^i\beta^j\beta^k}{\alpha}\left(-4c_4-2c_5 \right)\right. \\
	& \left. +c_5\alpha Q^j\beta^i+c_5\alpha Q^i\beta^j  \right].
	\end{split}
\end{align}
If we plug the aforementioned coefficients making null the velocities, we get the following primary constraint
\begin{align}
	\begin{split}
	\label{eq:HessianGTEGRsym0}
	{}^SC^{ij}={}^SP^{ij}-\frac{M_{pl}\sqrt{h}}{\alpha^2}&\left[b_2\alpha T^A{}_{k}{}^i\theta_A{}^k\beta^j  +b_2\alpha T^A{}_k{}^j\theta_A{}^k\beta^i  +\frac{\alpha}{2}Q^{jik}\beta_k\left(-b_2+c_2 \right)+\frac{\alpha}{2}Q^{ijk}\beta_k\left(-b_2+c_2 \right) \right. \\
	& \left.  +Q^j{}_{kl}\frac{\beta^i\beta^k\beta^l}{\alpha}\left(-2c_2-\frac{b_2}{2} \right) +Q^i{}_{kl}\frac{\beta^j\beta^k\beta^l}{\alpha}\left(-2c_2-\frac{b_2}{2} \right)+2c_4\alpha\bar{Q}^j\beta^i +2c_4\alpha\bar{Q}^i\beta^j\right. \\
	& \left. + \bar{Q}_k\frac{\beta^i\beta^j\beta^k}{\alpha}\left(-4c_4-b_2 \right) +\frac{b_2}{2}\alpha Q^j\beta^i+\frac{b_2}{2}\alpha Q^i\beta^j  \right]-{}^S\Pi^{ij} \approx 0.
	\end{split}
\end{align}

	There are also primary constraints associated with the induced metric, whose occurrence automatically yields no propagating spin-2 field. Thus, if such conditions are satisfied, the resulting theory cannot be understood as a gravity model, but rather as a different theory relying on a rank-2 tensor. The momentum reads:
 	\begin{align}
	\begin{split}
	\pi^{(ij)}=\frac{M_{pl}^2\sqrt{h}}{\alpha^3}&\left[-2c_1\alpha^2Q_0{}^{ij}-2c_3\alpha^2h^{ij}Q_0+(2c_3+c_5)h^{ij}Q_{000}\right. \\
	& \left. +2(2c_3+c_5)h^{ij}Q_{00}{}^k\beta_k+2(2c_3+c_5)h^{ij}Q_{0kl}\beta^k\beta^l \right. \\
	& \left. -c_5h^{ij}\alpha^2\bar{Q}_0-(2c_3+c_5)h^{ij}Q_{k00}\beta^k+2(2c_3+c_5)h^{ij} Q_{k0l}\beta^k\beta^l \right. \\
	& \left.+2c_1\alpha^2Q_k{}^{ij}\beta^k+c_2\alpha^2 Q^{ijk}\beta_k+c_2\alpha^2 Q^{jik}\beta_k+2c_3h^{ij}\alpha^2 Q_k\beta^k \right. \\
	& \left. +c_5h^{ij}\alpha^2\bar{Q}_k\beta^k-(2c_3+c_5)h^{ij}Q_{klm}\beta^k\beta^l\beta^m 
-b_2\alpha^2h^{ij}T^A{}_{0k}\theta_A{}^k-b_1\alpha^2 T^A{}_0{}^{(i}\theta_A{}^{j)} \right. \\ &\left. +b_1\alpha T^A{}_0{}^{(i}\beta^{j)}\xi_A-b_1T^A{}_{0k}\beta^i\beta^j\beta^k \right].
	\end{split}
	\end{align}
 For general and symmetric teleparallel theories, the following choice makes the spin-2 field nondynamical:
	\begin{align}
	b_1=b_2=c_1=c_3=c_5=0,
	\end{align}
	gives rise to the following 6 primary constraints
			\begin{align}
	\begin{split}
\label{eq:nogravity}
	{}^{\mathrm{NG}}C^{(ij)}=\pi^{(ij)}\approx 0,
	\end{split}
	\end{align} 
	where $\mathrm{NG}$ stands for ``No Gravity''. As stated in \cite{Dambrosio:2020wbi}, the imposition $c_1=c_3=c_5=0$ gives us primary constraints for symmetric teleparallelism.

 In the context of symmetric teleparallelism, primary constraints have not been derived, except for STEGR \cite{Dambrosio:2020wbi,DAmbrosio:2020nqu} and $f(Q)$ \cite{DAmbrosio:2023asf,Heisenberg:2023lru,Hu:2022anq,Tomonari:2023wcs}. In \cite{Dambrosio:2020wbi}, the analysis of the Hessian reveals that primary constraints occur either when $c_1=c_3=c_5=0$ or $c_4=-c_2+\frac{c_5^2}{4c_3}$, though the explicit expression of the primary constraints is not presented. However, it can be seen that $c_1=c_3=c_5$ gives primary constraints for the full symmetric part in symmetric teleparallel gravity. In symmetric teleparallel quadratic gravity, it is possible to get the explicit expression of six primary constraints, occurring if $c_1=c_3=c_5=0$. Reasonably, the condition $c_4=-c_2+\frac{c_5^2}{4c_3}$ gives rise to one primary constraint associated to the trace, but this scenario will not be further investigated here, since this sector is non-gravitational due to the absence of spin-2 fields. In metric teleparallelism the primary constraints related to the non-gravitational sector are trivial \cite{Blixt:2020ekl}, as discussed in Sec. \ref{sec:PCTEGR}.

\begin{figure}[!ht]
		\includegraphics[width=12cm, height=14cm]{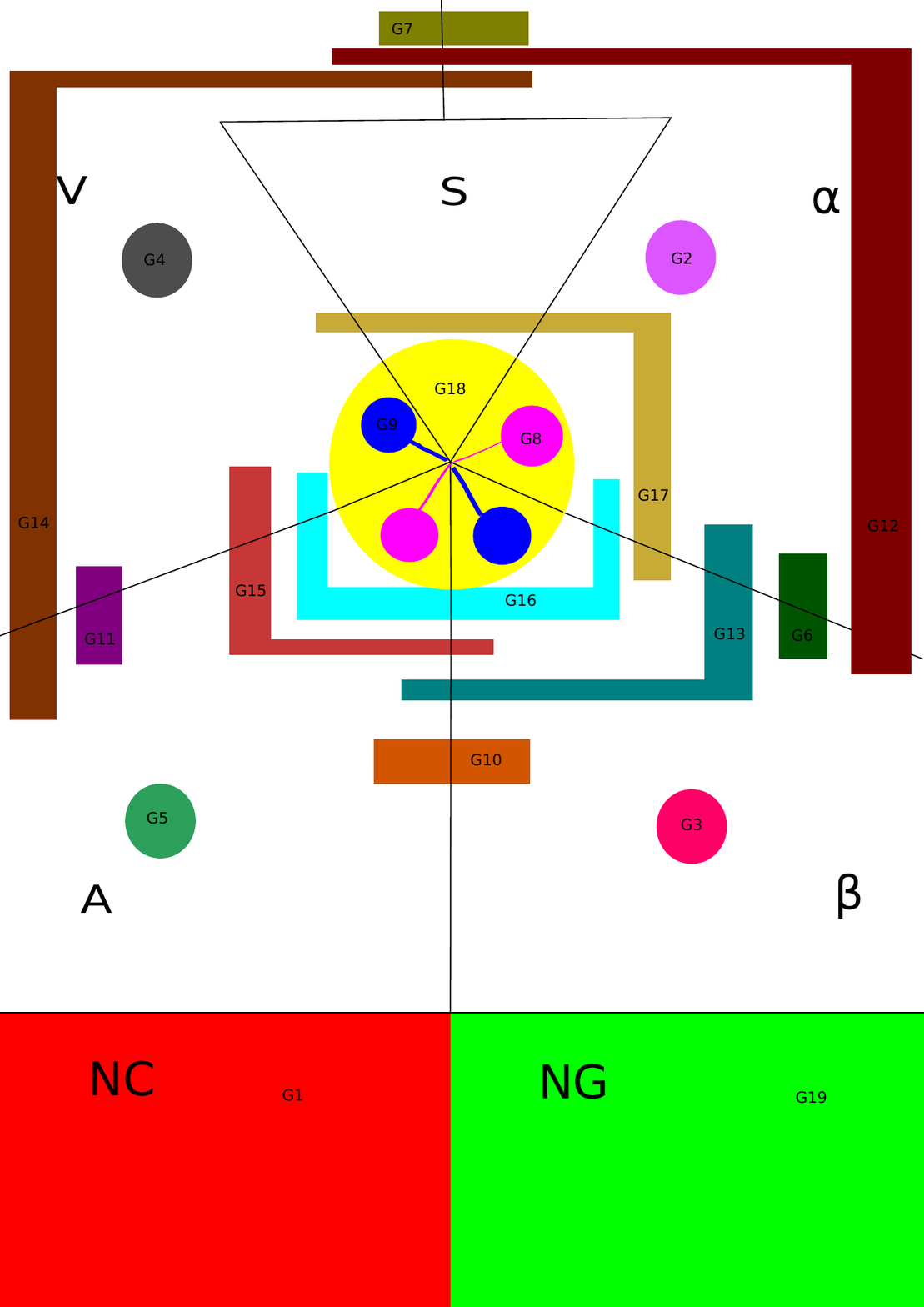}
  \includegraphics[width=5cm]{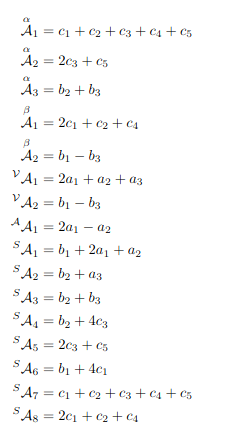}
		\caption{Visualization the constraints that G1-G19 theories satisfy, respectively. The lower left region (NC)  is the most generic theory with no constraint. The lower right corner (NG) stands for ``no gravity'' and is constrained such that the spin-2 field is non-propagating. The five corners, \emph{i.e.} $\alpha,\beta,\mathrm{V,A,S}$, represent the parameter spaces that have primary constraints for lapse, shift, vector, antisymmetric and symmetric, respectively. The combinations for coefficients that impose primary constraints are collected to the right.} 
		\centering
			\label{fig:class}
	\end{figure}

	In this study, we aim to present the conditions for primary constraints in general teleparallel quadratic gravity, from which it is possible to classify their different combinations and count the number of possibilities. Note that, as already demonstrated in the case of metric teleparallelism \cite{Blixt:2018znp} and symmetric teleparallelism \cite{Dambrosio:2020wbi}, often degeneracy can occur. Hence, certain combinations derived from combinatorial analysis, can represent trivial theories; on the other hand, a combination of specific primary constraints could impose an additional constraint. All possibilities are listed and displayed respectively in Table \ref{GeneralTable} and in Fig. \ref{fig:class}.
	
	\begin{table}[ht!]
		\begin{tabular}{cccc}
			\hline Name & Theory & Constraints & \# free parameters \\ \hline
			G1	&	$\mathcal{A}_I \neq 0$ \  $\forall$ \ $I \ \in \{\alpha,\beta,\mathcal{V},\mathcal{A},\mathcal{S},\mathcal{T}\}$  & No constraints & 10 \\ \hline 
			G2	&	$\overset{\alpha}{\mathcal{A}}_1=\overset{\alpha}{\mathcal{A}}_2=\overset{\alpha}{\mathcal{A}}_3=0$ & $\overset{\alpha}{C}$ & 7 \\ \hline
			G3	&	$\overset{\beta}{\mathcal{A}}_1=\overset{\beta}{\mathcal{A}}_2=0$ & $\overset{\beta}{C}{}^i$ & 8 \\  \hline 
			G4	&	${}^{\mathcal{V}}{\mathcal{A}}_1={}^{\mathcal{V}}{\mathcal{A}}_2=0$ & ${}^{\mathcal{V}}C^i$ & 8 \\  \hline 
			G5	&	${}^{\mathcal{A}}{\mathcal{A}}_1=0$ & ${}^{\mathcal{A}}C^{[ji]}$ & 9  \\ \hline 
			G6	&	$\overset{\alpha}{\mathcal{A}}_1=\overset{\alpha}{\mathcal{A}}_2=\overset{\alpha}{\mathcal{A}}_3=\overset{\beta}{\mathcal{A}}_1=\overset{\beta}{\mathcal{A}}_2=0$ & $\overset{\alpha}{C}, \ \overset{\beta}{C}{}^i$ & 5 \\ \hline 
			G7	&	$\overset{\alpha}{\mathcal{A}}_1=\overset{\alpha}{\mathcal{A}}_2=\overset{\alpha}{\mathcal{A}}_3={}^{\mathcal{V}}{\mathcal{A}}_1={}^{\mathcal{V}}{\mathcal{A}}_2=0$ & $\overset{\alpha}{C}, \ {}^{\mathcal{V}}C^i$ & 5 \\ \hline 
			G8	&	$\overset{\alpha}{\mathcal{A}}_1=\overset{\alpha}{\mathcal{A}}_2+2\overset{\alpha}{\mathcal{A}}_3={}^{\mathcal{A}}{\mathcal{A}}_1=0$ & $\overset{\alpha}{C}, \ {}^{\mathcal{A}}C^{[ji]}$ & 7 \\ \hline 
			G9	&	$\overset{\beta}{\mathcal{A}}_1=\overset{\beta}{\mathcal{A}}_2={}^{\mathcal{V}}\mathcal{A}_1=0$ & $\overset{\beta}{C}{}^i, \ {}^{\mathcal{V}}C^i$ & 7 \\ \hline
			G10	&	$\overset{\beta}{\mathcal{A}}_1=\overset{\beta}{\mathcal{A}}_2={}^{\mathcal{A}}\mathcal{A}_1=0$ & $\overset{\beta}{C}{}^i, \ {}^{\mathcal{A}}C^{[ji]}$ & 7 \\ \hline 
			G11	&	${}^{\mathcal{V}}{\mathcal{A}}_1={}^{\mathcal{V}}{\mathcal{A}}_2={}^{\mathcal{A}}\mathcal{A}_1=0$ & ${}^{\mathcal{V}}C^i, \ {}^{\mathcal{A}}C^{[ji]}$ & 7 \\ \hline 
			G12	&	$\overset{\alpha}{\mathcal{A}}_1=\overset{\alpha}{\mathcal{A}}_2=\overset{\alpha}{\mathcal{A}}_3=\overset{\beta}{\mathcal{A}}_1=\overset{\beta}{\mathcal{A}}_2={}^{\mathcal{V}}{\mathcal{A}}_1=0$ & $\overset{\alpha}{C}, \ \overset{\beta}{C}{}^i, \ {}^{\mathcal{V}}C^i$ & 4 \\ \hline 
			G13	&	$\overset{\alpha}{\mathcal{A}}_1=\overset{\alpha}{\mathcal{A}}_2=\overset{\alpha}{\mathcal{A}}_3=\overset{\beta}{\mathcal{A}}_1=\overset{\beta}{\mathcal{A}}_2={}^{\mathcal{A}}{\mathcal{A}}_1=0$ & $\overset{\alpha}{C}, \ \overset{\beta}{C}{}^i, \ {}^{\mathcal{A}}C^{[ji]}$ & 4 \\ \hline 
			G14	&	$\overset{\alpha}{\mathcal{A}}_1=\overset{\alpha}{\mathcal{A}}_2=\overset{\alpha}{\mathcal{A}}_3={}^{\mathcal{V}}\mathcal{A}_1={}^{\mathcal{V}}\mathcal{A}_2={}^{\mathcal{A}}{\mathcal{A}}_1=0$ & $\overset{\alpha}{C}, \ {}^{\mathcal{V}}C^i , \ {}^{\mathcal{A}}C^{[ji]}$ & 4 \\ \hline
			G15	&	$\overset{\beta}{\mathcal{A}}_1=\overset{\beta}{\mathcal{A}}_2={}^{\mathcal{V}}\mathcal{A}_1={}^{\mathcal{A}}\mathcal{A}_1=0$ & $\overset{\beta}{C}{}^i, \ {}^{\mathcal{V}}C^i , \ {}^{\mathcal{A}}C^{[ji]}$ & 6 \\ \hline 
			G16	&	$\overset{\alpha}{\mathcal{A}}_1=\overset{\alpha}{\mathcal{A}}_2=\overset{\alpha}{\mathcal{A}}_3={}^{\mathcal{V}}\mathcal{A}_1=\overset{\beta}{\mathcal{A}}_1={}^{\mathcal{V}}\mathcal{A}_2={}^{\mathcal{A}}{\mathcal{A}}_1=0$ & $\overset{\alpha}{C}, \ \overset{\beta}{C}{}^i , \ {}^{\mathcal{V}}C^i , \ {}^{\mathcal{A}}C^{[ji]}$ & 3 \\ \hline 
			G17	&	${}^{\mathcal{S}}\mathcal{A}_i=\overset{\alpha}{\mathcal{A}}_i=\overset{\beta}{\mathcal{A}}_i={}^{\mathcal{V}}\mathcal{A}_i=0$ & $ {}^{\mathrm{S}}C^{ij}, \ \overset{\alpha}{C}, \ \overset{\beta}{C}^i,  {}^{\mathcal{V}}C^i$ & 2 \\ \hline 
				G18		&	${}^{\mathcal{S}}\mathcal{A}_i=\overset{\alpha}{\mathcal{A}}_i=\overset{\beta}{\mathcal{A}}_i={}^{\mathcal{V}}\mathcal{A}_i={}^{\mathcal{A}}\mathcal{A}_i=0$ & $ {}^{\mathrm{S}}C^{ij}, \ \overset{\alpha}{C}, \ \overset{\beta}{C}^i,  {}^{\mathcal{V}}C^i,  {}^{\mathcal{A}}C^{[ji]}$ & 1 \\ \hline 
					G19	&	$b_1=b_2=c_1=c_3=c_5=0$ & $ {}^{\mathrm{NG}}C^{(ji)}$ & 5 \\ \hline 
		\end{tabular}
		\caption{Classification of general teleparallel quadratic theories based on primary constraints}
		\label{GeneralTable}
	\end{table}
	In the Table \ref{GeneralTable} we have introduced the index $i$ for brevity. It should be understood as running through all possible labels of the corresponding $\mathcal{A}$. For example one should view $\overset{\alpha}{\mathcal{A}}_i$ as taking any of the values $\overset{\alpha}{\mathcal{A}}_1,\overset{\alpha}{\mathcal{A}}_2,\overset{\alpha}{\mathcal{A}}_3$.

	\section{Primary constraints in the trinity of gravity and its nonlinear extensions}
	
	\label{sec:PCtrinity}
	
	Let us begin by recalling the values to which the coefficients $c_i$ must be set to restore GTEGR:
	\begin{align}
 \label{coeffGTEGR}
	&(a_1,a_2,a_3)=\left(\frac{1}{4},\frac{1}{2},-1 \right), \indent (b_1,b_2,b_3)=(-1,1,-1) \\
	&\mathrm{and} \indent (c_1,c_2,c_3,c_4,c_5)=\left(\frac{1}{4},-\frac{1}{2},-\frac{1}{4},0,\frac{1}{2} \right).
 \label{coeffGTEGR1}
	\end{align}
	Notice that these coefficients satisfy the conditions for the presence of primary constraints, except for  ${}^{\mathrm{NG}}C^{ij}$ (or their irreducible parts). The case of TEGR is obtained in the limit $b_1=b_2=b_3=c_1=c_2=c_3=c_4=c_5=0$, while to get STEGR we must set $b_1=b_2=b_3=a_1=a_2=a_3=0$. Consequently, we can obtain the primary constraints for the tinity of gravity by simply plugging these coefficients into the result obtained in the previous section. Note that we still have the following theory-independent constraints\footnote{The same contraints will be used throughout the rest of the paper}:
\begin{align}
\label{eq:PCTempTrinity}
{}^{P}C_\mu{}^0=P_\mu{}^0-\overset{\alpha}{\pi}\overset{\alpha}{S}_\mu+\overset{\beta}{\pi}{}_i\overset{\beta}{S}{}^i_\mu\approx 0,
\end{align}

 The GTEGR coefficients have a similar role for nonlinear extensions. 
    Let us then assume that extended GTEGR model can be recast as a second-order theory non-minimally coupled to a dynamical scalar field, namely
	\begin{align}
 \label{eq:f(G)}
	S_{f(G)}=\frac{1}{2}M^2_{pl}\int \mathrm{d}^4x \sqrt{-g}f(G)\equiv \frac{1}{2}M^2_{pl}\int \mathrm{d}^4x \sqrt{-g}\left(\phi G -V(\phi)\right),
	\end{align} 
where $G$ is defined as the argument of the action in Eq. \eqref{GTEGRction}, with the coefficients set as in Eqs. \eqref{coeffGTEGR} and \eqref{coeffGTEGR1}, namely
\begin{align}
	\begin{split}
	G \equiv & \frac{1}{4} T_{\alpha\mu\nu}T^{\alpha\mu\nu}+ \frac{1}{2} T_{\alpha\mu\nu}T^{\nu\mu\alpha} - T_\mu T^\mu  \\
	& \left. - Q_{\alpha\mu\nu}T^{\nu\alpha\mu} + Q_\mu T^\mu - \bar{Q}_\mu T^\mu\right. \\
	& + \frac{1}{4} Q_{\alpha\mu\nu}Q^{\alpha\mu\nu} - \frac{1}{2}Q_{\alpha\mu\nu}Q^{\mu\nu\alpha} - \frac{1}{4} Q_\mu Q^\mu +  \frac{1}{2} Q_\mu\bar{Q}^\mu.
	\end{split}
	\end{align}

This is not proved and we take it for granted, assuming that under conformal transformations modified GTEGR behaves like $f(R)$ \cite{Cognola:2007zu} and $f(Q)$ \cite{Heisenberg:2023lru} gravities. By imposing vanishing non-metricity or torsion, $f(T)$ and $f(Q)$ models are respectively restored. In all cases, we can immediately note the primary constraints yield
	\begin{align}
	\overset{\phi}{\pi}:=\frac{\partial L}{\partial \dot{\phi}}\approx 0.
	\end{align}
	It is worthwhile stressing the difference between $f(G)$ and $f(R)$ gravity. In the latter case, the velocities $\dot{\phi}$ appear because of the presence of second order derivatives in $R$, which can be cast in terms of time derivatives of $\phi$ prior integrating by parts. The same does not occur in the latter case, namely in $f(G)$, since $G$ does not contain higher order derivatives. The presence of the scalar field alters the primary constraints in the trinity of gravity, due to a re-scaling of the conjugate momenta. This introduces a relative factor of $\phi$ alongside other terms within the primary constraints (if present), correspondingly breaking the symmetry, while activating new degrees of freedom found in the nonlinear extensions to the trinity of gravity. In the following subsections we will list the primary constraints of the trinity of gravity as well as the primary constraints in their nonlinear extensions.

	\subsection{GTEGR and $f(G)$}
	
	\label{sec:PCGTEGR}

	Let us start by writing the GTEGR primary constraint related to lapse function $\alpha$:
	\begin{align}
	\begin{split}
	\label{eq:alphaPCGTEGR}
	\overset{\alpha}{C}=\overset{\alpha}{\pi}+\frac{M_{pl}\sqrt{h}}{2\alpha^2} \left(\bar{Q}_0-\bar{Q}_i\beta^i \right)\approx 0,
	\end{split}
	\end{align}	
	which coincides with the corresponding primary constraint in STEGR. By plugging the coefficients \eqref{coeffGTEGR} and \eqref{coeffGTEGR1} into the shift primary constraint \eqref{Cbeta}, we get
	\begin{align}
	\begin{split}
	\label{eq:betaPCGTEGR}
	\overset{\beta}{C}{}^m=\overset{\beta}{\pi}{}^m+\frac{M_{pl}\sqrt{h}}{2}&\left[\frac{1}{2\alpha^3}\left(Q_{i00}+Q_{ijk}\beta^j\beta^k-2Q_{i0j}\beta^j \right)+\frac{Q^m}{2\alpha} -\frac{T^m}{\alpha}  \right]\approx 0,
	\end{split}
	\end{align}
	which is equivalent to the corresponding primary constraint of STEGR, except for the presence of an additional torsion term. This result suggests that diffeomorphism invariance emerges in a more intricate way than the case of STEGR. Adopting the same procedure for the vector constraint yields
	\begin{align}
	\begin{split}
	\label{eq:VPCGTEGR}
	{}^{\mathcal{V}}C^i={}^{\mathcal{V}}P^i+\frac{M_{pl}\sqrt{h}}{\alpha}&\left[\theta^B{}_j\eta_{AB} T^{Aji} +\frac{\alpha}{2} Q^i-\frac{\alpha}{2}\bar{Q}^i \right]\approx 0.
	\end{split}
	\end{align}	
	Here we notice that the primary constraint is influenced by the mixed term and this gives rise to the presence of non-metricity. If one assume non-metricity to vanish, the corresponding primary constraints for TEGR can be recovered. The final primary constraint is given by the antisymmetric irreducible part and reads as:
	\begin{align}
	\begin{split}
	\label{eq:APCGTEGR}
	{}^{\mathcal{A}}C^{[ji]}={}^{\mathcal{A}}P^{[ji]}-\frac{\sqrt{h}M_{pl}}{4}&\left[2 \xi_A T^{Aij}+\alpha^2Q^j{}_0{}^i-\alpha^2Q^i{}_0{}^j- Q^{jik}\beta_k+ Q^{ijk}\beta_k\right]\approx 0.
	\end{split}
	\end{align}
	Also in this case, the above constraint contains non-metricity terms in addition to the expected torsion term. This means that both diffeomorphism and Lorentz invariance are realized in a more complicate way, if compared to canonical gravity, STEGR and TEGR. Finally, the symmetric constraints associated with the relation between metric and affine connection reduces to
\begin{align}
\begin{split}
\label{eq:PCSGTEGR}
{}^SC^{ij}={}^SP^{ij}-\frac{M_{pl}\sqrt{h}}{\alpha^2}&\left[\alpha T^A{}_{k}{}^i\theta_A{}^k\beta^j  +\alpha T^A{}_k{}^j\theta_A{}^k\beta^i  -\frac{3}{4}\alpha Q^{jik}\beta_k-\frac{3}{4}\alpha Q^{ijk}\beta_k   +Q^j{}_{kl}\frac{\beta^i\beta^k\beta^l}{2\alpha} +Q^i{}_{kl}\frac{\beta^j\beta^k\beta^l}{2\alpha}\right. \\
& \left. - \bar{Q}_k\frac{\beta^i\beta^j\beta^k}{\alpha} +\frac{1}{2}\alpha Q^j\beta^i+\frac{1}{2}\alpha Q^i\beta^j  \right]-{}^S\Pi^{ij} \approx 0.
\end{split}
\end{align}
	
	 Extending this result to the nonlinear extension of GTEGR is straightforward by recasting the $f(G)$ model in the Jordan frame, namely as a scalar tensor theory of the form presented in Eq. \eqref{eq:f(G)}. As shown in Refs. \cite{Blixt:2020ekl,Heisenberg:2023lru}, the equivalence between the two frames can be realized both in $f(T)$ and $f(Q)$ gravity. From this \emph{ansatz} it follows that the aforementioned primary constraints will be deformed by a factor $\phi$, which in turn carries an additional constraint. Thus the whole set of primary constraints in $f(G)$ gravity reads as:
 \begin{align}
     \overset{\phi}{\pi}\approx 0,
 \end{align}
 	\begin{align}
	\begin{split}
	\label{eq:alphaPCfG}
	\overset{\alpha}{C}_{f(G)}=\overset{\alpha}{\pi}+\phi\frac{M_{pl}\sqrt{h}}{2\alpha^2} \left(\bar{Q}_0-\bar{Q}_i\beta^i \right)\approx 0,
	\end{split}
	\end{align}	
		\begin{align}
	\begin{split}
	\label{eq:betaPCfG}
	\overset{\beta}{C}{}^m_{f(G)}=\overset{\beta}{\pi}{}^m+\phi\frac{M_{pl}\sqrt{h}}{2}&\left[\frac{1}{2\alpha^3}\left(Q_{i00}+Q_{ijk}\beta^j\beta^k-2Q_{i0j}\beta^j \right)+\frac{Q^m}{2\alpha} -\frac{T^m}{\alpha}  \right]\approx 0,
	\end{split}
	\end{align}
	\begin{align}
	\begin{split}
	\label{eq:VPCfG}
	{}^{\mathcal{V}}C^i_{f(G)}={}^{\mathcal{V}}\pi^i+\phi\frac{M_{pl}\sqrt{h}}{\alpha}&\left[\theta^B{}_j\eta_{AB} T^{Aji} +\frac{\alpha}{2} Q^i-\frac{\alpha}{2}\bar{Q}^i \right]\approx 0,
	\end{split}
	\end{align}	
	\begin{align}
	\begin{split}
	\label{eq:APCfG}
	{}^{\mathcal{A}}C^{[ji]}_{f(G)}={}^{\mathcal{A}}\pi^{[ji]}-\phi\frac{\sqrt{h}M_{pl}}{4}&\left[2 \xi_A T^{Aij}+\alpha^2Q^j{}_0{}^i-\alpha^2Q^i{}_0{}^j- Q^{jik}\beta_k+ Q^{ijk}\beta_k\right]\approx 0,
	\end{split}
	\end{align}

\begin{align}
\begin{split}
\label{eq:HessianGTEGRsym}
{}^SC^{ij}={}^SP^{ij}-\phi\frac{M_{pl}\sqrt{h}}{\alpha^2}&\left[\alpha T^A{}_{k}{}^i\theta_A{}^k\beta^j  +\alpha T^A{}_k{}^j\theta_A{}^k\beta^i  -\frac{3}{4}\alpha Q^{jik}\beta_k-\frac{3}{4}\alpha Q^{ijk}\beta_k   +Q^j{}_{kl}\frac{\beta^i\beta^k\beta^l}{2\alpha} +Q^i{}_{kl}\frac{\beta^j\beta^k\beta^l}{2\alpha}\right. \\
& \left. - \bar{Q}_k\frac{\beta^i\beta^j\beta^k}{\alpha} +\frac{1}{2}\alpha Q^j\beta^i+\frac{1}{2}\alpha Q^i\beta^j  \right]-{}^S\Pi^{ij} \approx 0.
\end{split}
\end{align}
One can conclude that calculating the Poisson brackets needs cumbersome computations and that the intricate realization of symmetries in GTEGR will be consequently deformed.

	\subsection{STEGR and $f(Q)$}
	\label{sec:PCSTEGR}
	The limit of STEGR can be obtained by assuming a vanishing torsion in the action\footnote{Note that one cannot simply take the primary constraints of GTEGR and set the torsion to zero, since they originate from a variation with respect to torsion.} \eqref{GTEGRction} and properly setting the values of the coefficients $a_i$, $b_i$, $c_i$. The vector and antisymmetric irreducible parts are pure tetrad contributions that trivially vanish ${}^{\mathcal{V}}\pi^i={}^{\mathcal{A}}\pi^{[ij]}=0$, as symmetric teleparallelism can be formulated completely in terms of the metric. The conjugate momenta with respect to lapse and shift are, in contrast to canonical gravity, dependent on non-metricity itself:
	\begin{align}
	\begin{split}
	\label{eq:alphaPCSTEGR}
	\overset{\alpha}{C}=\overset{\alpha}{\pi}+\frac{M_{pl}\sqrt{h}}{2\alpha^2} \left(\bar{Q}_0-\bar{Q}_i\beta^i \right)\approx 0,
	\end{split}
	\end{align}	
	\begin{align}
	\begin{split}
	\label{eq:betaPCSTEGR}
	\overset{\beta}{C}{}^m=\overset{\beta}{\pi}{}^m+\frac{M_{pl}\sqrt{h}}{2}&\left[\frac{1}{2\alpha^3}\left(Q_{i00}+Q_{ijk}\beta^j\beta^k-2Q_{i0j}\beta^j \right)+\frac{Q^m}{2\alpha}   \right]\approx 0.
	\end{split}
	\end{align}
 Notice that the former constraint turns out to be the same as in GTEGR. Finally, there are momenta with respect to the connection related to primary constraints. Apart from the four temporal constraints, mentioned at the beginning of this section, we also have the following twelve constraints occurring for any symmetric teleparallel theory
 \begin{align}
 \label{eq:CoincidentGauge}
 	{}^{\mathcal{V}}C^i={}^\mathcal{V}P^i\approx 0, \indent {}^{\mathcal{A}}C^{[ji]}={}^\mathcal{A}P^{[ji]}\approx 0, \indent {}^{\mathcal{S}}C^i={}^\mathrm{S}P^{ij}-{}^{\mathrm{S}}\Pi^{ij} \approx 0. 
\end{align}
This is consistent with the findings of Refs. \cite{BeltranJimenez:2022azb,Blixt:2023kyr}, though the expressions provided in the latter reference differ since they do not adopt the ADM-variables. In other words, the coincident gauge is always allowed in symmetric teleparallel theories, but the same does not hold in general teleparallel theories.
 
 The nonlinear extension $f(Q)$, can be dealt with by considering the action \eqref{eq:f(G)} and setting $b_i = c_i = 0$. Moreover, similarly to the case of $f(Q)$ it is convenient to consider the Einstein frame, so that a further constraint arises due to the presence of the scalar field $\phi$. The whole set of primary constraints in $f(Q)$ gravity is:
\begin{align}
    \overset{\phi}{C}_{f(Q)}=\overset{\phi}{\pi}\approx 0,
\end{align}
		\begin{align}
	\begin{split}
	\label{eq:alphaPCfQ}
	\overset{\alpha}{C}_{f(Q)}=\overset{\alpha}{\pi}+\phi\frac{M_{pl}\sqrt{h}}{2\alpha^2} \left(\bar{Q}_0-\bar{Q}_i\beta^i \right)\approx 0,
	\end{split}
	\end{align}	
 	\begin{align}
	\begin{split}
	\label{eq:betaPCfQ}
	\overset{\beta}{C}{}^m_{f(Q)}=\overset{\beta}{\pi}{}^m+\phi\frac{M_{pl}\sqrt{h}}{2}&\left[\frac{1}{2\alpha^3}\left(Q_{i00}+Q_{ijk}\beta^j\beta^k-2Q_{i0j}\beta^j \right)+\frac{Q^m}{2\alpha}   \right]\approx 0.
	\end{split}
	\end{align}
 Nevertheless, an alternative form of the primary constraints is presented in the literature \cite{Hu:2022anq,DAmbrosio:2023asf,Heisenberg:2023lru,Tomonari:2023wcs}, with a nontrivial expression for $\overset{\phi}{\pi}$, which is claimed to be obtained after an integration by parts. The references also made different conclusions regarding the number of propagating degrees of freedom, which suggests that the Hamiltonian structure for symmetric teleparallel theories needs to undergo a more extensive analysis in order to clarify disputing results.

	\subsection{TEGR and $f(T)$}
	\label{sec:PCTEGR}
	Let us finally consider the most well-studied theory within the trinity of gravity, TEGR. Enforcing non-metricity to vanish at the level of the action, it is straightforward to realize that there are no time derivatives in the lapse and shift sectors, thus ${}^{\alpha}\pi={}^{\beta}\pi_i=0$. However, as pointed out in \cite{Blixt:2020ekl}, the primary constraints for the antisymmetric and vector parts contain torsion, contrary to the case of canonical gravity. The primary constraints then read
	\begin{align}
	\begin{split}
	\label{eq:VPCTEGR}
	{}^{\mathcal{V}}C^i={}^{\mathcal{V}}\pi^i+\frac{M_{pl}\sqrt{h}}{\alpha}& \theta^B{}_j\eta_{AB} T^{Aji}  \approx 0,
	\end{split}
	\end{align}	
	and
	\begin{align}
	\begin{split}
	\label{eq:APCTEGR}
	{}^{\mathcal{A}}C^{[ji]}={}^{\mathcal{A}}\pi^{[ji]}-\frac{\sqrt{h}M_{pl}}{4}& 2 \xi_A T^{Aij} \approx 0.
	\end{split}
	\end{align}
	They both differ from the GTEGR case due to the absence of non-metricity. In the absence of non-metricity, Eq. \eqref{eq:HessianGTEGRsym} does not trivialize, no associated primary constraint can be found and $L^A{}_\mu$ plays the role of a tetrad. It is known that the affine connection can be split in two parts, associated with tetrad and spin connection \cite{Krssak:2018ywd}. On the other hand, one can always choose the so-called Weitzenb\"ock gauge, where the dynamics is contained into the tetrads \cite{Golovnev:2022rui,Blagojevic:2023fys}, which appear along with the primary constraints. Alternatively, the same can be realized by coordinate transformation in the tangent space \cite{Blixt:2022rpl}. Here we found a new perspective for the Hamiltonian analysis, by simply using the sixteen components of $L^\mu{}_\nu$ as our starting point, instead of the sixteen components of the tetrad, but assuming the Weitzenb\"ock gauge. Previous works in the covariant formulation include six additional degrees of freedom associated with the spin connection and, in some cases, even curvature is included in the action as Lagrange multipliers. Note that, in metric teleparallel gravity, it is still possible to get primary constraints in the symmetric sector. However, this kills the degrees of freedom of the spin-2 field, contrary to the requisite for having a gravitational theory \cite{BeltranJimenez:2019nns} (the explicit expressions can be found in \cite{Blixt:2020ekl} and references therein). 
	
	The most well-studied extended teleparallel theory is $f(T)$ and, although suffering from inconsistencies \cite{Blixt:2020ekl}, several studies about the Hamiltonian analysis reveals the existence of five degrees of freedom \cite{Blagojevic:2020dyq}. The existing literature in this context agrees upon the form of the primary constraints to be 
 \begin{align}
    \overset{\phi}{C}_{f(T)}=\overset{\phi}{\pi}\approx 0,
\end{align}
\begin{align}
	\begin{split}
	\label{eq:VPCfT}
	{}^{\mathcal{V}}C^i_{f(T)}={}^{\mathcal{V}}\pi^i+\phi\frac{M_{pl}\sqrt{h}}{\alpha}& \theta^B{}_j\eta_{AB} T^{Aji}  \approx 0,
	\end{split}
	\end{align}	
 \begin{align}
	\begin{split}
	\label{eq:APCfT}
	{}^{\mathcal{A}}C^{[ji]}_{f(T)}={}^{\mathcal{A}}\pi^{[ji]}-\phi\frac{\sqrt{h}M_{pl}}{4}& 2 \xi_A T^{Aij} \approx 0.
	\end{split}
	\end{align}	
	
	\section{Conditions for primary constraints in the irreducible representation of torsion and non-metricity}
	\label{sec:conditionsPC}

	Torsion and non-metricity can be decomposed in irreducible parts \cite{Bahamonde:2022kwg, McCrea:1992wa, herrera, Vassiliev:2001qa,Pasic:2017zwe}, one of which plays a crucial role in the one-parameter healthy class of metric teleparallel quadratic gravity\footnote{In \cite{Golovnev:2023yla}, the statement that other theories propagate ghosts was proven to be incorrect and from this point of view the harsh conclusions of the viability of those theories can be weakened.}\cite{Aldrovandi:2013wha,Blixt:2019mkt}. The most famous irreducible part of non-metricity is accounted for the Weyl vector $W_\mu$. To complete the irreducible decomposition of non-metricity, four terms are needed (including the Weyl vector), \emph{i.e.}: 
	\begin{align}
	W_\mu=\frac{1}{4}Q_\mu,
	\end{align}
	\begin{align}
	\Lambda_\mu=\frac{4}{9}\left(\bar{Q}_\mu-W_\mu \right),
	\end{align}
	\begin{align}
	\Omega_\lambda{}^{\mu\nu}=-\left[\epsilon^{\mu\nu\rho\sigma}Q_{\rho\sigma\lambda}+\epsilon^{\mu\nu\rho}{}_{\lambda} \left(\frac{3}{4}\Lambda_\rho-W_\rho \right) \right],
	\end{align}
	\begin{align}
	q_{\lambda\mu\nu}=Q_{(\lambda\mu\nu)}-g_{(\mu\nu}W_{\lambda)}-\frac{3}{4}g_{(\mu\nu}\Lambda_{\lambda)}.
	\end{align}
	
	For the torsion we have three terms, namely
	\begin{align}
	v_\mu=T^\nu{}_{\mu\nu}, \\
	a_\mu=\epsilon_{\mu\lambda\rho\nu}T^{\lambda\rho\nu}, \\
	t_{\lambda\mu\nu}=T_{\lambda\mu\nu}-\frac{2}{3}g_{\lambda[\nu}T_{\mu]}-\frac{1}{6}\epsilon_{\lambda\rho\mu\nu}a^\rho,
	\end{align}
	named the vector, axial and tensor parts, respectively. According to this representation, the GTEGR scalar $G$ can be recast as
	\begin{align}
	\label{eq:irr}
	\begin{split}
	k_1 t_{\alpha\mu\nu}t^{\alpha\mu\nu}+k_2 a_\mu a^\mu +k_3 v_\mu v^\mu +k_4 \epsilon^{\mu\nu \alpha\beta }t_{\lambda\mu\nu}\Omega^{\lambda}{}_{\alpha\beta} +k_5 v_\mu W^\mu +k_6 v_\mu \Lambda^\mu +k_7 q_{\alpha\mu\nu}q^{\alpha\mu\nu} \\
	+k_8 \Omega_\alpha{}^{\mu\nu} \Omega^\alpha{}_{\mu\nu} +k_9 \Lambda_\mu\Lambda^\mu +k_{10} W_\mu W^\mu+k_{11} W_\mu \Lambda^\mu.
	\end{split}
	\end{align}
	Most of the terms in the above expression can be written explicitly in terms of Weyl vector, torsion and non-metricity. In particular, we have:
	
	\begin{align}
	\begin{split}
	k_7q_{\lambda\mu\nu}q^{\lambda\mu\nu}=&k_7\left[Q_{(\lambda\mu\nu)}Q^{(\lambda\mu\nu)}-2Q_{(\lambda\mu\nu)}g^{(\mu\nu}W^{\lambda)}-\frac{3}{2}Q_{(\lambda\mu\nu)}g^{(\mu\nu}\Lambda^{\lambda)}+g_{(\mu\nu}W_{\lambda)}g^{(\mu\nu}W^{\lambda)} \right. \\
	& \left. +\frac{3}{2}g_{(\mu\nu}W_{\lambda)}g^{(\mu\nu}\Lambda^{\lambda)}+\frac{9}{16}g_{(\mu\nu}\Lambda_{\lambda)}g^{(\mu\nu}\Lambda^{\lambda)} \right] \\
	=&k_7\left[\frac{1}{2}Q_{\lambda\mu\nu}Q^{\mu\lambda\nu}+\frac{1}{2}Q_{\lambda\mu\nu}Q^{\lambda\mu\nu} +\frac{1}{36}\bar{Q}_{\lambda}Q^{\lambda} +\frac{11}{72}Q_\lambda Q^\lambda -\frac{1}{18}\bar{Q}_{\lambda}\bar{Q}^{\lambda}  \right],
	\end{split}
	\end{align}
	
	\begin{align}
	\begin{split}
	k_8\Omega_\lambda{}^{\mu\nu}\Omega^\lambda{}_{\mu\nu}=&\left[\epsilon^{\mu\nu\rho\sigma}Q_{\rho\sigma\lambda}+\epsilon^{\mu\nu\rho}{}_\lambda\left(\frac{3}{4}\Lambda_\rho-W_\rho \right)\right]\left[\epsilon_{\mu\nu\alpha\beta}Q^{\alpha\beta\lambda}+\epsilon_{\mu\nu\alpha}{}^\lambda\left(\frac{3}{4}\Lambda^\alpha-W^\alpha \right)\right]\\
	=&\epsilon^{\mu\nu\rho\sigma}Q_{\rho\sigma\lambda}\epsilon_{\mu\nu\alpha\beta}Q^{\alpha\beta\lambda}+\epsilon^{\mu\nu\rho\sigma}Q_{\rho\sigma\lambda}\epsilon_{\mu\nu\alpha}{}^\lambda\left(\frac{3}{4}\Lambda^\alpha-W^\alpha \right)\\
	& +\epsilon^{\mu\nu\rho}{}_\lambda\left(\frac{3}{4}\Lambda_\rho-W_\rho \right)\epsilon_{\mu\nu\alpha\beta}Q^{\alpha\beta\lambda}+\epsilon^{\mu\nu\rho}{}_\lambda\left(\frac{3}{4}\Lambda_\rho-W_\rho \right)\epsilon_{\mu\nu\alpha}{}^\lambda\left(\frac{3}{4}\Lambda^\alpha-W^\alpha \right)\\
	=&2 Q_{\alpha\beta\lambda}Q^{\alpha\beta\lambda}-2Q_{\beta\alpha\lambda}Q^{\alpha\beta\lambda} -\frac{2}{3}Q_\lambda Q^\lambda-\frac{2}{3}\bar{Q}_\lambda\bar{Q}^\lambda+\frac{4}{3}Q_\lambda\bar{Q}^\lambda ,
	\end{split}
	\end{align}
	
	\begin{align}
	\begin{split}
	k_9\Lambda_\mu \Lambda^\mu=&\frac{16}{81}\left(\bar{Q}_\mu-\frac{1}{4}Q_\mu\right)\left(\bar{Q}^\mu-\frac{1}{4}Q^\mu\right)=\frac{16}{81}\bar{Q}_\mu\bar{Q}^\mu-\frac{8}{81}Q_\mu\bar{Q}^\mu+\frac{1}{81}Q_\mu Q^\mu ,
	\end{split}
	\end{align}
	
	\begin{align}
	k_{10} W_\mu W^\mu = \frac{k_{10}}{16}Q_\mu Q^\mu,
	\end{align}

	\begin{align}
	k_{11}W_\mu \Lambda^\mu =\frac{1}{9}Q_\mu(\bar{Q}^\mu-\frac{1}{4}Q^\mu),
	\end{align}
	from which it is possible to get the following system of linear equations in the symmetric teleparallel case
	\begin{align}
	c_1 Q_{\lambda\mu\nu}Q^{\lambda\mu\nu}=Q_{\lambda\mu\nu}Q^{\lambda\mu\nu}\left(\frac{1}{2}k_7+2k_8 \right), \\
	c_2 Q_{\lambda\mu\nu}Q^{\mu\lambda\nu}= Q_{\lambda\mu\nu}Q^{\mu\lambda\nu}\left(\frac{1}{2}k_7-2k_8 \right), \\
	c_3Q_\mu Q^\mu=Q_\mu Q^\mu \left(-\frac{7}{72}k_7-\frac{2}{3}k_8+\frac{1}{81}k_9+\frac{1}{16}k_{10}-\frac{1}{36}k_{11} \right), \\
	c_4 \bar{Q}_\mu \bar{Q}^\mu =\bar{Q}_\mu \bar{Q}^\mu \left(-\frac{1}{18}k_7-\frac{2}{3}k_8+\frac{16}{81}k_9+\frac{1}{9}k_{11} \right), \\
	c_5 \bar{Q}_\mu Q^\mu = \bar{Q}_\mu Q^\mu \left(-\frac{2}{9}k_7+\frac{4}{3}k_8-\frac{8}{81}k_9 \right)  ,
	\end{align}
	whose solution is
	\begin{align}
	\begin{cases}
	k_7=c_1+c_2 \\
	k_8=\frac{c_1-c_2}{4} \\
	k_9 =\frac{9}{8}c_1-\frac{45}{8}c_2-\frac{81}{8}c_5 \\
	k_{10}  =16c_3+4c_4+10c_5 \\
	k_{11}	= 9c_2+9c_4+18c_5 
	\end{cases} \begin{cases}
	c_1=\frac{1}{2}k_7+2k_8 \\
	c_2=\frac{1}{2}k_7-2k_8 \\
	c_3=\frac{11}{72}k_7-\frac{2}{3}k_8+\frac{1}{81}k_9+\frac{1}{16}k_{10}-\frac{1}{36}k_{11} \\
	c_4=-\frac{1}{18}k_7-\frac{2}{3}k_8+\frac{16}{81}k_9+\frac{1}{9}k_{11 } \\
	c_5=-\frac{2}{9}k_7+\frac{4}{3}k_8-\frac{8}{81}k_9
	\end{cases}
	\end{align}
    If we substitute the STEGR-coefficients in the above solution, we obtain that the only non-vanishing coefficients $k_i$ are:
	\begin{align}
	\begin{cases}
	k_7=-\frac{1}{4} \\
	k_8=\frac{3}{16} \\
	k_9 =-\frac{63}{32} \\
	k_{10}  =1 \\
	k_{11}	= \frac{9}{2} \, ,
	\end{cases} 
	\end{align}
	
	so that conditions for primary constraints take the following form 
	\begin{align}
	\begin{cases}
	\overset{\alpha}{\mathcal{A}}_1=\frac{7}{8}k_7+\frac{1}{9}k_9+\frac{1}{16}k_{10}+\frac{1}{12}k_{11} \\
	\overset{\alpha}{\mathcal{A}}_2=\frac{1}{12}k_7-\frac{2}{27}k_9+\frac{1}{8}k_{10}-\frac{1}{18}k_{11} \\
	\overset{\beta}{\mathcal{A}}_1=\frac{13}{9}k_7+\frac{4}{3}k_8+\frac{16}{81}k_9+\frac{1}{9}k_{11}.
	\end{cases}
	\end{align}
	Note that the term $k_8\Omega_\lambda{}^{\mu\nu} \Omega^\lambda{}_{\mu\nu}$ is only involved in the condition for constraints related to shift, and not for lapse. However, the whole set of coefficients $c_i$ are involved in the primary constraint for lapse, in the standard action formulation for symmetric teleparallel theories. Furthermore, we also notice that $k_{10}W_\mu W^\mu$ is not involved in the condition for primary constraint related to the shift sector. However, this is not an improvement from the previous formulation, as neither $c_3$ nor $c_5$ determine any condition for this primary constraint. 
	
	In future works it would be interesting to check the further restrictions that can be obtained by demanding the Hamiltonian to be linear in lapse and shift.
	
	The case of metric teleparallelism is already known (see Ref. \cite{Blixt:2019ene} for details) and leads to the following solution for the coefficients $a_i$: 
	\begin{align}
	\begin{cases}
	a_1=\frac{1}{2}k_1-\frac{1}{18}k_2 \\
	a_2=\frac{1}{2}k_1+\frac{1}{9}k_2 \\
	a_3=k_3-\frac{1}{2}k_1
	\end{cases} \begin{cases}
	k_1=\frac{2}{3}(2a_1+a_2) \\
	k_2=6(a_2-a_1) \\
	k_3=\frac{1}{3}(2a_1+a_2)+a_3 \, ,
	\end{cases}
	\end{align}
	
	from which the conditions for primary constraints automatically follow
	\begin{align}
	\begin{cases}
	{}^{\mathcal{V}}\mathcal{A}_1=k_1+k_3\\
	{}^{\mathcal{A}}\mathcal{A}_1=\frac{1}{2}k_1-\frac{2}{9}k_2\\
	{}^{\mathcal{S}}\mathcal{A}_1=\frac{3}{2}k_1\\
	{}^{\mathcal{T}}\mathcal{A}_1=3k_3
	\end{cases}.
	\end{align}
	Finally, in order to evaluate the case of quadratic metric teleparallel theories of gravity, it is necessary to consider non-vanishing trace and symmetric sectors, thus imposing $k_1$, $k_3$ $\neq 0$, otherwise no propagating spin-2 field occurs and the physical relevance of the corresponding theory is consequently lost. Fixing one of them with the Planck mass and the ghost-free condition ${}^{\mathcal{V}}\mathcal{A}_1=0$, makes $k_2$ the only free parameter. The latter governs the axial part and, if set to $k_2=\frac{9}{4}k_1$, allows to restore TEGR, though the remaining parameters lead to a new physics that have been shown to suffer from the strong coupling problem \cite{BeltranJimenez:2019nns,Cheng:1988zg}.
	
	Similarly to the previous case, by considering the explicit expression of $k_i$ coefficients, namely
	
	\begin{align}
	\begin{split}
	k_4 \epsilon^{\mu\nu \alpha\beta }t_{\lambda\mu\nu}\Omega^{\lambda}{}_{\alpha\beta} =-12T_{\lambda\mu\nu}Q^{\mu\nu\lambda},
	\end{split}
	\end{align}

	\begin{align}
	\begin{split}
	k_5v_\mu W^\mu = \frac{1}{4}k_5 T_\mu Q^\mu,
	\end{split}
	\end{align}
	\begin{align}
	\begin{split}
	k_6v_\mu \Lambda^\mu=k_6 \frac{4}{9}T_\mu \left(\bar{Q}^\mu-\frac{1}{4}Q^\mu\right),
	\end{split}
	\end{align}
	we can write the following system of linear equations
	\begin{align}
	b_1Q_{\alpha\mu\nu}T^{\nu\alpha\mu}=-12k_4Q_{\alpha\mu\nu}T^{\nu\alpha\mu}, \\
	b_2Q_\mu T^\mu= Q_\mu T^\mu (\frac{1}{4}k_5-\frac{1}{9}k_6), \\
	b_3\bar{Q}_\mu T^\mu=\frac{4}{9}k_6,
	\end{align}
	whose solution yields
	\begin{align}
	\begin{cases}
	k_4=-\frac{1}{12}b_1 \\
	k_5=4b_2+\frac{4}{9}b_3 \\
	k_6=\frac{9}{4}b_3
	\end{cases} \begin{cases}
	b_1=-12k_4 \\
	b_2=\frac{1}{4}k_5-\frac{1}{9}k_6 \\
	b_3=\frac{4}{9}k_6 \, .
	\end{cases}
	\end{align}
	Finally, the symmetric constraint related to GTEGR read
	\begin{align}
		\begin{cases}
		{}^{\mathrm{S}}\mathcal{A}_1=-12k_4+\frac{3}{2}k_1 \\
		{}^{\mathrm{S}}\mathcal{A}_2=\frac{1}{4}k_5-\frac{1}{9}k_6+k_3-\frac{1}{2}k_1 \\
		{}^{\mathrm{S}}\mathcal{A}_3=\frac{1}{4}k_5+\frac{1}{3}k_6 \\
		{}^{\mathrm{S}}\mathcal{A}_4=\frac{1}{4}k_5-\frac{1}{9}k_6+\frac{11}{18}k_7-\frac{8}{3}k_8+\frac{4}{81}k_9+\frac{1}{4}k_{10}-\frac{1}{9}k_{11} \\
		{}^{\mathrm{S}}\mathcal{A}_5=\frac{1}{12}k_7-\frac{2}{27}k_9+\frac{1}{8}k_{10}-\frac{1}{18}k_{11}  \\
		{}^{\mathrm{S}}\mathcal{A}_6=-12k_4+2k_7+8k_8 \\
		{}^{\mathrm{S}}\mathcal{A}_7=\frac{7}{8}k_7+\frac{1}{9}k_9+\frac{1}{16}k_{10}+\frac{1}{12}k_{11} \\
		{}^{\mathrm{S}}\mathcal{A}_8=\frac{13}{9}k_7+\frac{4}{3}k_8+\frac{16}{81}k_9+\frac{1}{9}k_{11} 
		\end{cases}
\end{align} 
	From the results within the scope of this article, we cannot see any advantage of writing torsion and non-metricity into irreducible parts. However, as noted for quadratic symmetric teleparallel gravity, we found $k_8\Omega_\lambda{}^{\mu\nu}\Omega^\lambda{}_{\mu\nu}$ to play a special role regarding the primary constraints for lapse and shift, such as axial torsion plays a special role in metric teleparallel quadratic gravity.

	\section{Shifted algebra among constraints}
	\label{sec:Shifted}
	\noindent It is well-known that the trinity of gravity (including GTEGR) gives rise to the same field equations as GR. The degrees of freedom are, hence, two (or four in phase-space) and there must exist enough symmetry to constrain the extra introduced components of the tetrad and spin connections. The first point to stress is that the spin connection is purely gauge, which has support in the literature i) for the case of metric teleparallelism \cite{Golovnev:2023yla,Blixt:2022rpl,Blixt:2019mkt,Blagojevic:2023fys,Golovnev:2023qll}, ii) for the case of symmetric teleparallelism \cite{Blixt:2023kyr,Golovnev:2023qll}, and iii) but not for the case of general teleparallelism. Thus, in metric and symmetric teleparallel theories we are allowed to choose the so-called coincident gauge \cite{Bahamonde:2022zgj, Zhao:2021zab, Hohmann:2021ast} where the affine connection drops out and the focus shifts only on the 10 (or 20 in phase space) degrees of freedom of the metric. This means that the associated primary constraints are expected to be of the first class, removing the degrees of freedom of the affine connection, whereas the case of GTEGR works differently. As a matter of facts, spin connection is pure gauge in any teleparallel theory, but the affine connection can be made to vanish only in symmetric teleparallelism. In other words the Weitzenb\"ock gauge is always available, but the coincident gauge is available only for symmetric teleparallelism \cite{BeltranJimenez:2022azb}
 
    For the canonical Hamiltonian analysis, however, the output is influenced by boundary terms \cite{Corichi:2023ery}. This in turn affects the algebra among constraints and, consequently, the way to restore the symmetries of GR. We name this feature ``shifted algebra among constraints'' and this section is dedicated to shading light in this framework by comparing both the three corners of the trinity of gravity and GTEGR.
	
	\subsection{The case of canonical gravity}
	
	In the canonical case, namely with the Einstein-Hilbert action as the starting point, the spin connection is not present. Since the theory is fully invariant under Lorentz transformations, it is conventionally formulated in terms of metric rather than tetrad \cite{Arnowitt:1962hi}, especially because residing to the tetrad formulation makes the analysis more cumbersome. Nevertheless, the latter approach can be consistently pursued at the same level of the metric one, as demonstrated in \cite{Peldan:1993hi,Hinterbichler:2012cn}. The tetrad formulation includes the antisymmetric part of the conjugate momenta, which introduces primary constraints simply providing
	\begin{align}
 \label{eq:PCtetradCanonical}
	\pi^{[\mu\nu]}\approx 0.
	\end{align}
	This can also be written in the canonical form, with opposite indices with respect to those of the tetrad, if one adopts the $\mathcal{V},\mathcal{A}$ part of the irreducible decomposition with respect to the rotation group $\mathcal{O}(3)$. However, for simplicity, in this section we follow the standard formulation based on the metric. The primary constraints restrict the additional six components arising when passing from the metric to the tetrad formulation. By calculating the Poisson brackets with the Hamiltonian, it turns out that the well-known Lorentz invariance of GR follows from the Lorentz algebra provided by the Poisson brackets.
	
	Let us start by considering, in the metric formulation, the trivial primary constraints associated with lapse and shift:
	\begin{align}
	\overset{\alpha}{\pi}\approx 0 , \indent 	\overset{\beta}{\pi}{}^i \approx 0.
	\end{align}
	They are related to diffeomorphism invariance and with the intrinsic nature of lapse and shift, which are purely gauge degrees of freedom. However, them being first class only reduces the number of degrees of freedom to 6 (or 12 in phase space). This happens because diffeomorphism invariance in canonical gravity does not only give rise to primary constraints, but also to secondary constraints and, for this reason, diffeomorphism invariance can be thought as ``hitting twice'' \cite{Golovnev:2022rui}. The latter consideration finally brings the counting down to 2 degrees of freedom (or 4 in phase-space). The secondary constraints are realized by the observation that the final expression for the Hamiltonian turns out to be linear in lapse and shift. In the following we will demonstrate how the counting of degrees of freedom goes similarly in the trinity of gravity. The symmetries are the same, but their generators are shifted, as it can be realized by the more intricate expressions for the primary constraints provided in Sec. \ref{sec:PCtrinity}.
	
	\subsection{The case of TEGR}
	The case of TEGR is particularly interesting in our formulation, since the metric was only introduced through non-metricity. Thus, in the metric teleparallel limit, we only have degrees of freedom related to $L^\mu{}_\nu$. It turns out that the latter takes the same role as the tetrad in the Weitzenb\"ock gauge and torsion can also be rewritten in terms of tetrad and the spin-connection derivatives \cite{Krssak:2024xeh}. Consequently, one can choose 16 components of the tetrad plus 6 components of Lorentz matrices as the starting point. The Hamiltonian then reveals that all Lorentz matrix components are subjected to primary constraints \cite{Golovnev:2021omn,Blagojevic:2023fys} and choosing the Weitzenb\"ock gauge is physically fully consistent \cite{Blixt:2022rpl,Krssak:2024xeh}. In this approach the starting point is the covariant formulation, as the coincident gauge is inconsistent with general teleparallelism (torsion would vanish and the geometry would reduce to symmetric teleparallelism). However, instead of considering tetrad plus Lorentz matrices (or spin connection), one can also start from $L^\mu{}_\nu$, which structurally appears in the same way as the tetrad appears in the Weitzenb\"ock gauge. This means that $L^\mu{}_\nu$ can be associated with the ``Lorentz gauge-invariant variables in metric teleparallel theories'' \cite{Blixt:2022rpl}.

 We found primary constraints for all general teleparallel theories associated with the temporal part of $L^\mu{}_\nu$ (see Eq. \eqref{eq:TempLConstraint}). In the Weitzenb\"ock analysis, this corresponds to the four primary constraints associated with lapse and shift \cite{Blixt:2020ekl}. Thus, the sixteen degrees of freedom coming from $L^\mu{}_\nu$ or $\tetrad$ reduce to twelve, if these are of first class. Additionally, there are the vector and antisymmetric constraints given by Eqs. \eqref{eq:VPCTEGR} and \eqref{eq:APCTEGR}\footnote{In the case of TEGR and its nonlinear extension $f(T)$-gravity, it is possible to combine the 3 vector primary constraints with the 3 antisymmetric primary constraints, to get 6 Lorentz constraints \cite{Maluf:2013gaa,Li:2011rn}.}. They are not independent of torsion, which makes the calculations of the Poisson brackets among constraints more difficult, compared to canonical gravity (Eq. \eqref{eq:PCtetradCanonical}). However, these computations have been performed (see \cite{Ferraro:2016wht}) and it can be shown that they form the Lorentz algebra. Therefore, though the Lorentz symmetry is realized by the Hamiltonian analysis, the generators are shifted. The remaining six degrees of freedom coincide with those of canonical gravity evaluated at the level of primary constraints, namely without considering secondary constraints from the Hamiltonian. Following the evaluation of these secondary constraints in TEGR \cite{Blagojevic:2000qs,Ferraro:2016wht}, the final conclusion is two degrees of freedom (or four in phase-space), which is consistent with GR. An interesting note is that there are no primary constraints associated with ${}^S{P}{}^{ij}$, contrarily to the cases of GTEGR and STEGR. This lack is due to the fact that the momentum ${}^S{P}{}^{ij}$ in metric teleparallelism takes the role of momentum for the induced metric. If this momentum was subjected to primary constraints, the spin-2 field would become non-dynamical, which is inconsistent with the properties we expect for a gravitational theory and, certainly, not the case of GR. 
	
	\label{sec:ShiftedTEGR}

	\subsection{The case of STEGR}
	
	\label{sec:ShiftedSTEGR}
	
	For STEGR introducing tetrad fields is not needed since, like canonical gravity, the metric is enough and the antisymmetric part of the conjugate momenta vanishes when adopting the tetrad formalism. Furthermore, it can be seen that there exist primary constraints associated with each component of $L^\mu{}_\nu$, which is consistent with \cite{Blixt:2023kyr} and with the claim that the coincident gauge is valid for any symmetric teleparallel theory \cite{BeltranJimenez:2022azb}. STEGR differs from canonical gravity by the presence of (linear) time derivatives of lapse and shift. Thus, their conjugate momenta do not vanish, but rather are proportional to non-metricity (not containing time derivatives). This allows to recover the primary constraints given by Eqs. \eqref{eq:alphaPCSTEGR} and \eqref{eq:betaPCSTEGR}. 
	
	Comparing this to canonical gravity, the primary constraints related to lapse and shift are shifted, similarly to how the vector and antisymmetric constraints were shifted in the case of TEGR. In most cases, the evolution of these primary constraints is not considered, but they are instead referred to as gauge fixing. However, it is important to avoid any gauge fixing in the Hamiltonian analysis, as pointed out in \cite{Golovnev:2022rui}. This is due to the fact that the gauge fixing of lapse and shift leads to missing control to the ``hitting twice'' of diffeomorphism invariance. 
	
	Before providing more details on this topic, some comments about the state of art for the Hamiltonian analysis for symmetric teleparallel gravity are in order. Only recently, progresses on this topic have been pursued \cite{DAmbrosio:2020nqu,Dambrosio:2020wbi,Hu:2022anq,Tomonari:2023wcs,DAmbrosio:2023asf,Guzman:2023oyl}, and the preservation of these constraints in time have never been investigated. In fact, the earliest article \cite{DAmbrosio:2020nqu} made an integration by parts and found the correct boundary term to supplement in order to make it identical to canonical gravity\footnote{Note that in canonical gravity something similar is done to the Einstein--Hilbert action \cite{Arnowitt:1962hi}, with the aim to neglect higher order time derivatives.}. In \cite{Guzman:2023oyl} the same boundary term is added, but then the authors discarded a term that is mistakenly identified as a boundary:
	\begin{align*}
	\alpha \sqrt{h} D_i(Q^i-\bar{Q}{}^i)=\alpha \partial_i\left( \sqrt{h} (Q^i-\bar{Q}{}^i)\right),
	\end{align*}
	obviously failing to be a boundary due to the presence of lapse. In Refs. \cite{DAmbrosio:2020nqu,Guzman:2023oyl}, the interesting effect coming from the boundary term, which makes the difference between STEGR and canonical gravity, is avoided by integration by parts (in particular involving time derivatives). In \cite{Tomonari:2023wcs} it is stated the proposition: \textit{Surface terms do change canonical momentum variables but do not change the symplectic structure.} Thus, the boundary term that was added in \cite{DAmbrosio:2020nqu} is not incorrect, but in our opinion less interesting, since it simply leads to canonical gravity. In future works it would be interesting to perform the full Hamiltonian analysis of STEGR without 
	adding this kind of boundary terms. This will reasonably shift the algebra among constraints similarly to the case of TEGR.
	
	Out of the 10 free components occurring in the metric (20 if one includes the conjugate momenta), the aforementioned primary constraints are not enough to reduce the degrees of freedom to two (or four in phase space). The missing ingredients are the secondary constraints, conventionally called ``Hamiltonian'' and ``momenta'' constraints. They are realized by the Hamiltonian that turns out to be linear in lapse and shift, and together with the primary constraints for lapse and shift, there are eight constraints which can be proven to be of first class. This reduces the number of dynamical degrees of freedom to two (or four in phase space), as expected. In \cite{DAmbrosio:2020nqu,Hu:2022anq,Tomonari:2023wcs} it is demonstrated that the Hamiltonian, along with the Poisson brackets among the Hamiltonian and the momenta constraints, form the ADM algebra, proving they are indeed of first class.

	\subsection{The case of GTEGR}
	
	\label{sec:ShiftedGTEGR}
	
With the list of primary constraints, we can now sketch the counting of degrees of freedom in GTEGR. We know that the latter is a gravitational theory including diffeomorphism and Lorentz invariance and that the number of degrees of freedom should be two (or four in phase-space). Therefore, it is reasonable to assume that all constraints are of first class, so that the count of degrees of freedom in the phase-space is no longer needed. The metric and $L^\mu{}_\nu$ have 10+16 degrees of freedom, four of which are related to the temporal part of $L^\mu{}_\nu$, like in any general teleparallel theory. The vector and antisymmetric constraints remove six more degrees of freedom (also here associated to $L^\mu{}_\nu$), so that the count drops to 10+6, where ten are linked to the metric and six to $L^\mu{}_\nu$. The final six degrees of freedom related to $L^\mu{}_\nu$ are fully constrained by the symmetric constraints ${}^SP^{ij}$. Thus, if a given model carries the aforementioned primary constraints, then the connection does not contain any dynamical degree of freedom, with the consequence that the corresponding gravitational theory can be always framed within symmetric teleparallel theories. On the other hand, the lapse and momentum constraints remove four additional degrees of freedom from the ten related to the metric. Therefore, we end up having only six metric degrees of freedom, as in canonical gravity. The last four constraints, which are needed to drop the degrees of freedom count to the known number two, are expected to come from the secondary (Hamiltonian and momentum) constraints, though further analyses in this direction are needed to investigate the nature of such constraints. 	

More specifically, in GTEGR time derivatives appear linearly on both lapse and shift, as well as the antisymmetric part of $L^\mu{}_\nu$. The primary constraints are given by Eqs. \eqref{eq:PCTempTrinity} and Eq. \eqref{eq:alphaPCGTEGR}-\eqref{eq:PCSGTEGR} and they are all different from canonical gravity due to the presence of torsion and non-metricity and also contain additional primary constraints related to $L^\mu{}{}_\nu$. So far, no previous work dealing with the Hamiltonian analysis of general teleparallel theories has been done. However, all primary constraints of GTEGR should be preserved in time and it is also expected that the Hamiltonian reveals linearity in lapse and shift. Starting with the aforementioned 10+16 components of metric and $L^\mu{}_\nu$, one can see that there are sixteen primary constraints in GTEGR associated with $L^\mu{}_\nu$ (see Eq. \eqref{eq:PCTempTrinity} and Eqs. \eqref{eq:VPCGTEGR}-\eqref{eq:PCSGTEGR}). Then we essentially end up with the starting point of canonical gravity, where one needs to find constraints for the ten metric components. The primary constraints related to lapse and shift are reported in Eqs. \eqref{eq:alphaPCGTEGR}-\eqref{eq:betaPCGTEGR}, whereas we expect that secondary constraints can be found from the Hamiltonian, similarly to canonical gravity. However, the primary constraints are more than those occurring in canonical gravity and most of them contains both torsion and non-metricity. Notice that GTEGR is an alternative formulation that only differ from GR by a boundary term, meaning that the algebra among constraints will eventually provide the same results as Einstein's theory, \emph{i.e.} two dynamical degrees of freedom. Nevertheless, analyzing how the algebra closes can be of extreme interest for future studies, especially considering that the primary constraint contains both non-metricity and torsion, meaning that the latter are both needed to realize diffeomorphism and Lorentz symmetry, or rather the combined group $GL(4,\mathbb{R})$. It remains to see how different the Hamiltonian and momenta constraints are with respect to the canonical case, though we expect that the Poisson brackets among constraints will realize $GL(4,\mathbb{R})$ in a way that is much less straightforward than the cases of TEGR and STEGR, because of the complex form taken by Hamiltonian and momenta constraints.
	
	\section{Conclusions and outlook}
	
	\label{sec:Conclusions}
	
We have derived all primary constraints for general teleparallel quadratic theories of gravity. They have been shown to be consistent with previous works \cite{Blagojevic:2000qs,Maluf:2013gaa,Li:2011rn,Ferraro:2016wht,Blixt:2018znp,Blixt:2020ekl,Dambrosio:2020wbi}. The most remarkable insight from the primary constraints is that they contain both torsion and non-metricity, which is a feature also enjoyed by in GTEGR. Since general teleparallel quadratic gravity is equivalent to GR (modulo a boundary term), diffeomorphism and Lorentz invariance of both type I and II (following the conventions of \cite{Blixt:2022rpl,Blixt:2023kyr}) have to be realized. Moreover, the primary constraints are ``shifted'' by involving torsion and non-metricity, with the consequence that the Poisson brackets among constraints vanish on-shell in a much less trivial way. 
	
	We also note that, while the Hamiltonian analysis for metric teleparallel theories has been widely studied (see \cite{Blixt:2020ekl} and references therein), the case of symmetric teleparallel gravity has gotten attention recently (see \cite{Heisenberg:2023lru} and references therein). Furthermore, in STEGR \cite{DAmbrosio:2020nqu,Guzman:2023oyl}, an integration by parts essentially yielding the case of canonical gravity has been pursued in Ref.~\cite{Arnowitt:1962hi}, whereas the study of the nonlinear extension exhibited strong disagreements \cite{Heisenberg:2023lru,Hu:2022anq,DAmbrosio:2023asf,Tomonari:2023wcs}. More precisely, the disagreement in the number of degrees of freedom in $f(Q)$ can be compared with the recent result from perturbation theory \cite{Gomes:2023tur,Heisenberg:2023wgk,Heisenberg:2023tho} revealing 7 degrees of freedom, among which at least 1 ghost degree of freedom. Perturbations around backgrounds revealing less degrees of freedom will instead suffer from strong coupling \cite{Gomes:2023tur}.

	We have classified general teleparallel quadratic theories based on the (non)presence of primary constraints. In total we find 19 distinct teleparallel quadratic theories (see Table \ref{GeneralTable} and Fig. \ref{fig:class}) with all of them having up to 10 free parameters. Among these, we find 1 non-gravitational theory, though it is expected that a few more can be found. The other 18 theories have at least 1 free parameters. However, considering that GTEGR does not manifest any free parameters, it is expected that conditions for secondary constraints will introduce more branches in the tree of general teleparallel theories, fixing the final parameter. If restricting the geometry to metric teleparallelism, one gets 9 classes of theories with 0-2 free parameters, all containing the Hamiltonian and momenta constraints (secondary constraints ensuring diffeomorphism symmetry) \cite{Blixt:2018znp,Guzman:2020kgh}. For the case of symmetric teleparallel quadratic gravity, conditions for secondary constraints are unknown, but a similar classification of theories was done in \cite{Dambrosio:2020wbi}. It is easy to see that our results are consistent with metric and symmetric teleparallel limits.
		
	Alternatively, general teleparallel quadratic theories can be written in terms of irreducible components of torsion and non-metricity, as shown in Eq. \eqref{eq:irr}. We found that the $k_8 \Omega_\alpha{}^{\mu\nu}\Omega^\alpha{}_{\mu\nu} $-term does not affect conditions for primary constraints related to lapse. Furthermore, as already known from the literature, the axial part of torsion (\emph{i.e.} $k_2 a_\mu a^\mu$) can be associated with the 1-parameter family of ghost-free metric teleparallel quadratic gravity \cite{Aldrovandi:2013wha,Blixt:2019ene}. Our study also introduced a smoother way to perform the Hamiltonian analysis in covariant metric teleparallel theories, namely using $L^\mu{}_\nu$ instead of tetrad plus spin connection, which is equivalent to adopt the Weitzenb\"ock gauge.
	
	Our results open up several directions towards further investigating the Hamiltonian analysis of general teleparallel theories, which in turn can be considered to determine the viability of the given model and the number of degrees of freedom\footnote{This may efficiently be done for the 19 different theories, by extending the method developed in \cite{Barker:2024grr}.}, as well as for the application to quantum cosmology \cite{Bajardi:2021tul, Bajardi:2023vcc, Miranda:2021oig, Bajardi:2020osh, Hu:2022anq}. In the trinity of gravity, GTEGR stands out as the most promising candidate to improve the notion of energy and entropy in GR \cite{Gomes:2022vrc} and it would be interesting to further investigate this through a full-fledged Hamiltonian analysis. Moreover, as stated in Sec. \ref{sec:Shifted}, another future perspective is to get the full picture about the way in which the Poisson brackets among constraints vanish in the trinity of gravity (which we only know for the case of canonical gravity and TEGR, but not for STEGR and GTEGR). Finally, one may also investigate if TEGR, STEGR or GTEGR models have an advantageous formulation for numerical relativity as pointed out in \cite{Capozziello:2021pcg,DAmbrosio:2020nqu,Guzman:2023oyl}.
	
	\begin{acknowledgements}
	    D.B. is grateful for helpful discussions with Alexey Golovnev, Débora Aguiar Gomes, María-José Guzmán, Sofia Vidal Guzmán, Damianos Iosifidis, Laur Järv, and Laxmipriya Pati. The authors acknowledge the "Istituto Nazionale di Fisica Nucleare" (INFN), Sezione di Napoli, iniziativa specifica GINGER. 
	\end{acknowledgements}

	\bibliographystyle{ieeetr}
	\bibliography{references.bib}

\begin{thebibliography}{100}

\bibitem{Will:2014kxa}
C.~M. Will, ``{The Confrontation between General Relativity and Experiment},''
  {\em Living Rev. Rel.}, vol.~17, p.~4, 2014.

\bibitem{Goroff:1985th}
M.~H. Goroff and A.~Sagnotti, ``{The Ultraviolet Behavior of Einstein
  Gravity},'' {\em Nucl. Phys. B}, vol.~266, pp.~709--736, 1986.

\bibitem{Frieman:2008sn}
J.~Frieman, M.~Turner, and D.~Huterer, ``{Dark Energy and the Accelerating
  Universe},'' {\em Ann. Rev. Astron. Astrophys.}, vol.~46, pp.~385--432, 2008.

\bibitem{DeWitt:1967yk}
B.~S. DeWitt, ``{Quantum Theory of Gravity. 1. The Canonical Theory},'' {\em
  Phys. Rev.}, vol.~160, pp.~1113--1148, 1967.

\bibitem{Barack:2018yly}
L.~Barack {\em et~al.}, ``{Black holes, gravitational waves and fundamental
  physics: a roadmap},'' {\em Class. Quant. Grav.}, vol.~36, no.~14, p.~143001,
  2019.

\bibitem{Hehl:1994ue}
F.~W. Hehl, J.~D. McCrea, E.~W. Mielke, and Y.~Ne'eman, ``{Metric affine gauge
  theory of gravity: Field equations, Noether identities, world spinors, and
  breaking of dilation invariance},'' {\em Phys. Rept.}, vol.~258, pp.~1--171,
  1995.

\bibitem{Capozziello:2003tk}
S.~Capozziello, S.~Carloni, and A.~Troisi, ``{Quintessence without scalar
  fields},'' {\em Recent Res. Dev. Astron. Astrophys.}, vol.~1, p.~625, 2003.

\bibitem{Nojiri:2010wj}
S.~Nojiri and S.~D. Odintsov, ``{Unified cosmic history in modified gravity:
  from F(R) theory to Lorentz non-invariant models},'' {\em Phys. Rept.},
  vol.~505, pp.~59--144, 2011.

\bibitem{Bamba:2012cp}
K.~Bamba, S.~Capozziello, S.~Nojiri, and S.~D. Odintsov, ``{Dark energy
  cosmology: the equivalent description via different theoretical models and
  cosmography tests},'' {\em Astrophys. Space Sci.}, vol.~342, pp.~155--228,
  2012.

\bibitem{Sanders:2002pf}
R.~H. Sanders and S.~S. McGaugh, ``{Modified Newtonian dynamics as an
  alternative to dark matter},'' {\em Ann. Rev. Astron. Astrophys.}, vol.~40,
  pp.~263--317, 2002.

\bibitem{Copeland:2006wr}
E.~J. Copeland, M.~Sami, and S.~Tsujikawa, ``{Dynamics of dark energy},'' {\em
  Int. J. Mod. Phys. D}, vol.~15, pp.~1753--1936, 2006.

\bibitem{Capozziello:2019cav}
S.~Capozziello, R.~D'Agostino, and O.~Luongo, ``{Extended Gravity
  Cosmography},'' {\em Int. J. Mod. Phys. D}, vol.~28, no.~10, p.~1930016,
  2019.

\bibitem{Nojiri:2017ncd}
S.~Nojiri, S.~D. Odintsov, and V.~K. Oikonomou, ``{Modified Gravity Theories on
  a Nutshell: Inflation, Bounce and Late-time Evolution},'' {\em Phys. Rept.},
  vol.~692, pp.~1--104, 2017.

\bibitem{Odintsov:2023weg}
S.~D. Odintsov, V.~K. Oikonomou, I.~Giannakoudi, F.~P. Fronimos, and E.~C.
  Lymperiadou, ``{Recent Advances in Inflation},'' {\em Symmetry}, vol.~15,
  no.~9, p.~1701, 2023.

\bibitem{Stelle:1976gc}
K.~S. Stelle, ``{Renormalization of Higher Derivative Quantum Gravity},'' {\em
  Phys. Rev. D}, vol.~16, pp.~953--969, 1977.

\bibitem{Bajardi:2020mdp}
F.~Bajardi, S.~Capozziello, and D.~Vernieri, ``{Non-local curvature and
  Gauss\textendash{}Bonnet cosmologies by Noether symmetries},'' {\em Eur.
  Phys. J. Plus}, vol.~135, no.~12, p.~942, 2020.

\bibitem{Bajardi:2021hya}
F.~Bajardi, D.~Vernieri, and S.~Capozziello, ``{Exact solutions in
  higher-dimensional Lovelock and AdS$_{5}$ Chern-Simons gravity},'' {\em
  JCAP}, vol.~11, no.~11, p.~057, 2021.

\bibitem{Halliwell:1986ja}
J.~J. Halliwell, ``{Scalar Fields in Cosmology with an Exponential
  Potential},'' {\em Phys. Lett. B}, vol.~185, p.~341, 1987.

\bibitem{Uzan:1999ch}
J.-P. Uzan, ``{Cosmological scaling solutions of nonminimally coupled scalar
  fields},'' {\em Phys. Rev. D}, vol.~59, p.~123510, 1999.

\bibitem{Clifton:2011jh}
T.~Clifton, P.~G. Ferreira, A.~Padilla, and C.~Skordis, ``{Modified Gravity and
  Cosmology},'' {\em Phys. Rept.}, vol.~513, pp.~1--189, 2012.

\bibitem{Urban:2020lfk}
Z.~Urban, F.~Bajardi, and S.~Capozziello, ``{The
  Noether\textendash{}Bessel-Hagen symmetry approach for dynamical systems},''
  {\em Int. J. Geom. Meth. Mod. Phys.}, vol.~17, no.~14, p.~2050215, 2020.

\bibitem{Sotiriou:2008rp}
T.~P. Sotiriou and V.~Faraoni, ``{f(R) Theories Of Gravity},'' {\em Rev. Mod.
  Phys.}, vol.~82, pp.~451--497, 2010.

\bibitem{DeFelice:2010aj}
A.~De~Felice and S.~Tsujikawa, ``{f(R) theories},'' {\em Living Rev. Rel.},
  vol.~13, p.~3, 2010.

\bibitem{Capozziello:2011et}
S.~Capozziello and M.~De~Laurentis, ``{Extended Theories of Gravity},'' {\em
  Phys. Rept.}, vol.~509, pp.~167--321, 2011.

\bibitem{Starobinsky:2007hu}
A.~A. Starobinsky, ``{Disappearing cosmological constant in f(R) gravity},''
  {\em JETP Lett.}, vol.~86, pp.~157--163, 2007.

\bibitem{Nojiri:2006gh}
S.~Nojiri and S.~D. Odintsov, ``{Modified f(R) gravity consistent with
  realistic cosmology: From matter dominated epoch to dark energy universe},''
  {\em Phys. Rev. D}, vol.~74, p.~086005, 2006.

\bibitem{Bajardi:2022ocw}
F.~Bajardi, R.~D'Agostino, M.~Benetti, V.~De~Falco, and S.~Capozziello,
  ``{Early and late time cosmology: the f(R) gravity perspective},'' {\em Eur.
  Phys. J. Plus}, vol.~137, no.~11, p.~1239, 2022.

\bibitem{Capozziello:2021goa}
S.~Capozziello {\em et~al.}, ``{Constraining Theories of Gravity by GINGER
  experiment},'' {\em Eur. Phys. J. Plus}, vol.~136, no.~4, p.~394, 2021.
\newblock [Erratum: Eur.Phys.J.Plus 136, 563 (2021)].

\bibitem{Capozziello:2002rd}
S.~Capozziello, ``{Curvature quintessence},'' {\em Int. J. Mod. Phys. D},
  vol.~11, pp.~483--492, 2002.

\bibitem{Starobinsky:1980te}
A.~A. Starobinsky, ``{A New Type of Isotropic Cosmological Models Without
  Singularity},'' {\em Phys. Lett. B}, vol.~91, pp.~99--102, 1980.

\bibitem{Poplawski:2010kb}
N.~J. Pop\l{}awski, ``{Cosmology with torsion: An alternative to cosmic
  inflation},'' {\em Phys. Lett. B}, vol.~694, pp.~181--185, 2010.
\newblock [Erratum: Phys.Lett.B 701, 672--672 (2011)].

\bibitem{Cabral:2019gzh}
F.~Cabral, F.~S.~N. Lobo, and D.~Rubiera-Garcia,
  ``{Einstein\textendash{}Cartan\textendash{}Dirac gravity with $U(1)$ symmetry
  breaking},'' {\em Eur. Phys. J. C}, vol.~79, no.~12, p.~1023, 2019.

\bibitem{Arcos:2004tzt}
H.~I. Arcos and J.~G. Pereira, ``{Torsion gravity: A Reappraisal},'' {\em Int.
  J. Mod. Phys. D}, vol.~13, pp.~2193--2240, 2004.

\bibitem{Casadio:2021zai}
R.~Casadio, I.~Kuntz, and G.~Paci, ``{Quantum fields in teleparallel gravity:
  renormalization at one-loop},'' {\em Eur. Phys. J. C}, vol.~82, no.~3,
  p.~186, 2022.

\bibitem{Krssak:2015lba}
M.~Kr\v{s}\v{s}\'ak, ``{Holographic Renormalization in Teleparallel Gravity},''
  {\em Eur. Phys. J. C}, vol.~77, no.~1, p.~44, 2017.

\bibitem{Maluf:2013gaa}
J.~W. Maluf, ``{The teleparallel equivalent of general relativity},'' {\em
  Annalen Phys.}, vol.~525, pp.~339--357, 2013.

\bibitem{Bahamonde:2015zma}
S.~Bahamonde, C.~G. B\"ohmer, and M.~Wright, ``{Modified teleparallel theories
  of gravity},'' {\em Phys. Rev. D}, vol.~92, no.~10, p.~104042, 2015.

\bibitem{Xu:2012jf}
C.~Xu, E.~N. Saridakis, and G.~Leon, ``{Phase-Space analysis of Teleparallel
  Dark Energy},'' {\em JCAP}, vol.~07, p.~005, 2012.

\bibitem{Krssak:2018ywd}
M.~Krssak, R.~J. van~den Hoogen, J.~G. Pereira, C.~G. B\"ohmer, and A.~A.
  Coley, ``{Teleparallel theories of gravity: illuminating a fully invariant
  approach},'' {\em Class. Quant. Grav.}, vol.~36, no.~18, p.~183001, 2019.

\bibitem{Obukhov:2002tm}
Y.~N. Obukhov and J.~G. Pereira, ``{Metric affine approach to teleparallel
  gravity},'' {\em Phys. Rev. D}, vol.~67, p.~044016, 2003.

\bibitem{Geng:2011ka}
C.-Q. Geng, C.-C. Lee, and E.~N. Saridakis, ``{Observational Constraints on
  Teleparallel Dark Energy},'' {\em JCAP}, vol.~01, p.~002, 2012.

\bibitem{Bajardi:2021tul}
F.~Bajardi and S.~Capozziello, ``{Noether symmetries and quantum cosmology in
  extended teleparallel gravity},'' {\em Int. J. Geom. Meth. Mod. Phys.},
  vol.~18, no.~supp01, p.~2140002, 2021.

\bibitem{Cai:2015emx}
Y.-F. Cai, S.~Capozziello, M.~De~Laurentis, and E.~N. Saridakis, ``{f(T)
  teleparallel gravity and cosmology},'' {\em Rept. Prog. Phys.}, vol.~79,
  no.~10, p.~106901, 2016.

\bibitem{Li:2010cg}
B.~Li, T.~P. Sotiriou, and J.~D. Barrow, ``{$f(T)$ gravity and local Lorentz
  invariance},'' {\em Phys. Rev. D}, vol.~83, p.~064035, 2011.

\bibitem{Ferraro:2006jd}
R.~Ferraro and F.~Fiorini, ``{Modified teleparallel gravity: Inflation without
  inflaton},'' {\em Phys. Rev. D}, vol.~75, p.~084031, 2007.

\bibitem{Wu:2010xk}
P.~Wu and H.~W. Yu, ``{The dynamical behavior of $f(T)$ theory},'' {\em Phys.
  Lett. B}, vol.~692, pp.~176--179, 2010.

\bibitem{Caruana:2020szx}
M.~Caruana, G.~Farrugia, and J.~Levi~Said, ``{Cosmological bouncing solutions
  in $f(T,B)$ gravity},'' {\em Eur. Phys. J. C}, vol.~80, no.~7, p.~640, 2020.

\bibitem{Bahamonde:2016grb}
S.~Bahamonde and S.~Capozziello, ``{Noether Symmetry Approach in $f(T,B)$
  teleparallel cosmology},'' {\em Eur. Phys. J. C}, vol.~77, no.~2, p.~107,
  2017.

\bibitem{Finch:2018gkh}
A.~Finch and J.~L. Said, ``{Galactic Rotation Dynamics in f(T) gravity},'' {\em
  Eur. Phys. J. C}, vol.~78, no.~7, p.~560, 2018.

\bibitem{Aljaf:2022fbk}
M.~Aljaf, E.~Elizalde, M.~Khurshudyan, K.~Myrzakulov, and A.~Zhadyranova,
  ``{Solving the $H_{0}$ tension in f(T) gravity through Bayesian machine
  learning},'' {\em Eur. Phys. J. C}, vol.~82, no.~12, p.~1130, 2022.

\bibitem{Bamba:2010wb}
K.~Bamba, C.-Q. Geng, C.-C. Lee, and L.-W. Luo, ``{Equation of state for dark
  energy in $f(T)$ gravity},'' {\em JCAP}, vol.~01, p.~021, 2011.

\bibitem{Anagnostopoulos:2022gej}
F.~K. Anagnostopoulos, V.~Gakis, E.~N. Saridakis, and S.~Basilakos, ``{New
  models and big bang nucleosynthesis constraints in f(Q) gravity},'' {\em Eur.
  Phys. J. C}, vol.~83, no.~1, p.~58, 2023.

\bibitem{Capozziello:2022tvv}
S.~Capozziello and M.~Shokri, ``{Slow-roll inflation in f(Q) non-metric
  gravity},'' {\em Phys. Dark Univ.}, vol.~37, p.~101113, 2022.

\bibitem{Bajardi:2020fxh}
F.~Bajardi, D.~Vernieri, and S.~Capozziello, ``{Bouncing Cosmology in f(Q)
  Symmetric Teleparallel Gravity},'' {\em Eur. Phys. J. Plus}, vol.~135,
  no.~11, p.~912, 2020.

\bibitem{Banerjee:2021mqk}
A.~Banerjee, A.~Pradhan, T.~Tangphati, and F.~Rahaman, ``{Wormhole geometries
  in $f(Q)$ gravity and the energy conditions},'' {\em Eur. Phys. J. C},
  vol.~81, no.~11, p.~1031, 2021.

\bibitem{DAgostino:2022tdk}
R.~D'Agostino and R.~C. Nunes, ``{Forecasting constraints on deviations from
  general relativity in f(Q) gravity with standard sirens},'' {\em Phys. Rev.
  D}, vol.~106, no.~12, p.~124053, 2022.

\bibitem{Gomes:2023tur}
D.~A. Gomes, J.~Beltr\'an~Jim\'enez, A.~J. Cano, and T.~S. Koivisto,
  ``{Pathological Character of Modifications to Coincident General Relativity:
  Cosmological Strong Coupling and Ghosts in f(Q) Theories},'' {\em Phys. Rev.
  Lett.}, vol.~132, no.~14, p.~141401, 2024.

\bibitem{Heisenberg:2023wgk}
L.~Heisenberg, M.~Hohmann, and S.~Kuhn, ``{Cosmological teleparallel
  perturbations},'' {\em JCAP}, vol.~03, p.~063, 2024.

\bibitem{Hayashi:1979qx}
K.~Hayashi and T.~Shirafuji, ``{New general relativity.},'' {\em Phys. Rev. D},
  vol.~19, pp.~3524--3553, 1979.
\newblock [Addendum: Phys.Rev.D 24, 3312--3314 (1982)].

\bibitem{Iorio:2012cm}
L.~Iorio and E.~N. Saridakis, ``{Solar system constraints on f(T) gravity},''
  {\em Mon. Not. Roy. Astron. Soc.}, vol.~427, p.~1555, 2012.

\bibitem{BeltranJimenez:2019nns}
J.~Beltr\'an~Jim\'enez and K.~F. Dialektopoulos, ``{Non-Linear Obstructions for
  Consistent New General Relativity},'' {\em JCAP}, vol.~01, p.~018, 2020.

\bibitem{Cheng:1988zg}
W.-H. Cheng, D.-C. Chern, and J.~M. Nester, ``{Canonical Analysis of the One
  Parameter Teleparallel Theory},'' {\em Phys. Rev. D}, vol.~38,
  pp.~2656--2658, 1988.

\bibitem{Golovnev:2023jnc}
A.~Golovnev, A.~N. Semenova, and V.~P. Vandeev, ``{Conformal transformations
  and cosmological perturbations in New General Relativity},'' {\em JCAP},
  vol.~04, p.~064, 2024.

\bibitem{BeltranJimenez:2019odq}
J.~Beltr\'an~Jim\'enez, L.~Heisenberg, D.~Iosifidis, A.~Jim\'enez-Cano, and
  T.~S. Koivisto, ``{General teleparallel quadratic gravity},'' {\em Phys.
  Lett. B}, vol.~805, p.~135422, 2020.

\bibitem{Blagojevic:2002du}
M.~Blagojevic, {\em {Gravitation and gauge symmetries}}.
\newblock 2002.

\bibitem{Blagojevic:1988pp}
M.~Blagojevic and M.~Vasilic, ``{Asymptotic Symmetry and Conserved Quantities
  in the Poincare Gauge Theory of Gravity},'' {\em Class. Quant. Grav.},
  vol.~5, pp.~1241--1257, 1988.

\bibitem{Blagojevic:2013xpa}
M.~Blagojevi\'c and F.~W. Hehl, eds., {\em {Gauge Theories of Gravitation}: {A
  Reader with Commentaries}}.
\newblock Singapore: World Scientific, 2013.

\bibitem{Blagojevic:2000qs}
M.~Blagojevic and I.~A. Nikolic, ``{Hamiltonian structure of the teleparallel
  formulation of GR},'' {\em Phys. Rev. D}, vol.~62, p.~024021, 2000.

\bibitem{Obukhov:1996ka}
Y.~N. Obukhov, E.~J. Vlachynsky, W.~Esser, and F.~W. Hehl, ``{Effective
  Einstein theory from metric affine gravity models via irreducible
  decompositions},'' {\em Phys. Rev. D}, vol.~56, pp.~7769--7778, 1997.

\bibitem{Pasic:2017zwe}
V.~Pasic, E.~Barakovic, and N.~Okicic, ``{A New Representation of the Field
  Equations of Quadratic Metric-Affine Gravity},'' 5 2017.

\bibitem{Pasic:2005qr}
V.~Pasic and D.~Vassiliev, ``{PP-waves with torsion and metric-affine
  gravity},'' {\em Class. Quant. Grav.}, vol.~22, pp.~3961--3976, 2005.

\bibitem{Vassiliev:2001qa}
D.~Vassiliev, ``{Pseudoinstantons in metric affine field theory},'' {\em Gen.
  Rel. Grav.}, vol.~34, pp.~1239--1265, 2002.

\bibitem{Vassiliev:2003dk}
D.~Vassiliev, ``{Quadratic metric affine gravity},'' {\em Annalen Phys.},
  vol.~517, pp.~231--252, 2005.

\bibitem{Gomes:2022vrc}
D.~A. Gomes, J.~Beltr\'an~Jim\'enez, and T.~S. Koivisto, ``{Energy and entropy
  in the geometrical trinity of gravity},'' {\em Phys. Rev. D}, vol.~107,
  no.~2, p.~024044, 2023.

\bibitem{Arnowitt:1962hi}
R.~L. Arnowitt, S.~Deser, and C.~W. Misner, ``{The Dynamics of general
  relativity},'' {\em Gen. Rel. Grav.}, vol.~40, pp.~1997--2027, 2008.

\bibitem{DeWitt:1967ub}
B.~S. DeWitt, ``{Quantum Theory of Gravity. 2. The Manifestly Covariant
  Theory},'' {\em Phys. Rev.}, vol.~162, pp.~1195--1239, 1967.

\bibitem{Wheeler:1957mu}
J.~A. Wheeler, ``{On the Nature of quantum geometrodynamics},'' {\em Annals
  Phys.}, vol.~2, pp.~604--614, 1957.

\bibitem{Vilenkin:1988yd}
A.~Vilenkin, ``{The Interpretation of the Wave Function of the Universe},''
  {\em Phys. Rev. D}, vol.~39, p.~1116, 1989.

\bibitem{Hawking:1983hj}
S.~W. Hawking, ``{The Quantum State of the Universe},'' {\em Nucl. Phys. B},
  vol.~239, p.~257, 1984.

\bibitem{Vilenkin:1982de}
A.~Vilenkin, ``{Creation of Universes from Nothing},'' {\em Phys. Lett. B},
  vol.~117, pp.~25--28, 1982.

\bibitem{Vilenkin:1984wp}
A.~Vilenkin, ``{Quantum Creation of Universes},'' {\em Phys. Rev. D}, vol.~30,
  pp.~509--511, 1984.

\bibitem{Bousso:2011up}
R.~Bousso and L.~Susskind, ``{The Multiverse Interpretation of Quantum
  Mechanics},'' {\em Phys. Rev. D}, vol.~85, p.~045007, 2012.

\bibitem{Capozziello:2011hj}
S.~Capozziello, V.~F. Cardone, H.~Farajollahi, and A.~Ravanpak, ``{Cosmography
  in f(T)-gravity},'' {\em Phys. Rev. D}, vol.~84, p.~043527, 2011.

\bibitem{Boehmer:2011gw}
C.~G. Boehmer, A.~Mussa, and N.~Tamanini, ``{Existence of relativistic stars in
  f(T) gravity},'' {\em Class. Quant. Grav.}, vol.~28, p.~245020, 2011.

\bibitem{Sotiriou:2009xt}
T.~P. Sotiriou, ``{f(R) gravity, torsion and non-metricity},'' {\em Class.
  Quant. Grav.}, vol.~26, p.~152001, 2009.

\bibitem{BeltranJimenez:2019esp}
J.~Beltr\'an~Jim\'enez, L.~Heisenberg, and T.~S. Koivisto, ``{The Geometrical
  Trinity of Gravity},'' {\em Universe}, vol.~5, no.~7, p.~173, 2019.

\bibitem{BeltranJimenez:2019tme}
J.~Beltr\'an~Jim\'enez, L.~Heisenberg, T.~S. Koivisto, and S.~Pekar,
  ``{Cosmology in $f(Q)$ geometry},'' {\em Phys. Rev. D}, vol.~101, no.~10,
  p.~103507, 2020.

\bibitem{Dialektopoulos:2019mtr}
K.~F. Dialektopoulos, T.~S. Koivisto, and S.~Capozziello, ``{Noether symmetries
  in Symmetric Teleparallel Cosmology},'' {\em Eur. Phys. J. C}, vol.~79,
  no.~7, p.~606, 2019.

\bibitem{JimenezCano:2021rlu}
A.~Jim\'enez~Cano, {\em {Metric-affine Gauge theories of gravity. Foundations
  and new insights}}.
\newblock PhD thesis, Granada U., Theor. Phys. Astrophys., 2021.

\bibitem{Blagojevic:2023fys}
M.~Blagojevi\'c and J.~M. Nester, ``{From the Lorentz invariant to the coframe
  form of f(T) gravity},'' {\em Phys. Rev. D}, vol.~109, no.~6, p.~064034,
  2024.

\bibitem{Blixt:2019mkt}
D.~Blixt, M.~Hohmann, and C.~Pfeifer, ``{On the gauge fixing in the Hamiltonian
  analysis of general teleparallel theories},'' {\em Universe}, vol.~5, no.~6,
  p.~143, 2019.

\bibitem{Blixt:2022rpl}
D.~Blixt, R.~Ferraro, A.~Golovnev, and M.-J. Guzm\'an, ``{Lorentz
  gauge-invariant variables in torsion-based theories of gravity},'' {\em Phys.
  Rev. D}, vol.~105, no.~8, p.~084029, 2022.

\bibitem{Blixt:2023kyr}
D.~Blixt, A.~Golovnev, M.-J. Guzman, and R.~Maksyutov, ``{Geometry and
  covariance of symmetric teleparallel theories of gravity},'' {\em Phys. Rev.
  D}, vol.~109, no.~4, p.~044061, 2024.

\bibitem{Golovnev:2021omn}
A.~Golovnev and M.-J. Guzman, ``{Lorentz symmetries and primary constraints in
  covariant teleparallel gravity},'' {\em Phys. Rev. D}, vol.~104, no.~12,
  p.~124074, 2021.

\bibitem{BeltranJimenez:2022azb}
J.~Beltr\'an~Jim\'enez and T.~S. Koivisto, ``{Lost in translation: The Abelian
  affine connection (in the coincident gauge)},'' {\em Int. J. Geom. Meth. Mod.
  Phys.}, vol.~19, no.~07, p.~2250108, 2022.

\bibitem{Blixt:2020ekl}
D.~Blixt, M.-J. Guzm\'an, M.~Hohmann, and C.~Pfeifer, ``{Review of the
  Hamiltonian analysis in teleparallel gravity},'' {\em Int. J. Geom. Meth.
  Mod. Phys.}, vol.~18, no.~supp01, p.~2130005, 2021.

\bibitem{Blagojevic:2000xd}
M.~Blagojevic, ``{Hamiltonian structure and gauge symmetries of Poincare gauge
  theory},'' {\em Annalen Phys.}, vol.~513, pp.~367--391, 2001.

\bibitem{Blixt:2018znp}
D.~Blixt, M.~Hohmann, and C.~Pfeifer, ``{Hamiltonian and primary constraints of
  new general relativity},'' {\em Phys. Rev. D}, vol.~99, no.~8, p.~084025,
  2019.

\bibitem{Pati:2022nwi}
L.~Pati, D.~Blixt, and M.-J. Guzman, ``{Hamilton\textquoteright{}s equations in
  the covariant teleparallel equivalent of general relativity},'' {\em Phys.
  Rev. D}, vol.~107, no.~4, p.~044071, 2023.

\bibitem{Dambrosio:2020wbi}
F.~D'ambrosio and L.~Heisenberg, ``{Classification of primary constraints of
  quadratic non-metricity theories of gravity},'' {\em JHEP}, vol.~02, p.~170,
  2021.

\bibitem{DAmbrosio:2020nqu}
F.~D'Ambrosio, M.~Garg, L.~Heisenberg, and S.~Zentarra, ``{ADM formulation and
  Hamiltonian analysis of Coincident General Relativity},'' 7 2020.

\bibitem{DAmbrosio:2023asf}
F.~D'Ambrosio, L.~Heisenberg, and S.~Zentarra, ``{Hamiltonian Analysis of
  f(Q)$f(\mathbb {Q})$ Gravity and the Failure of the
  Dirac\textendash{}Bergmann Algorithm for Teleparallel Theories of Gravity},''
  {\em Fortsch. Phys.}, vol.~71, no.~12, p.~2300185, 2023.

\bibitem{Heisenberg:2023lru}
L.~Heisenberg, ``{Review on f(Q) gravity},'' {\em Phys. Rept.}, vol.~1066,
  pp.~1--78, 2024.

\bibitem{Hu:2022anq}
K.~Hu, T.~Katsuragawa, and T.~Qiu, ``{ADM formulation and Hamiltonian analysis
  of f(Q) gravity},'' {\em Phys. Rev. D}, vol.~106, no.~4, p.~044025, 2022.

\bibitem{Tomonari:2023wcs}
K.~Tomonari and S.~Bahamonde, ``{Dirac\textendash{}Bergmann analysis and
  degrees of freedom of coincident f(Q)-gravity},'' {\em Eur. Phys. J. C},
  vol.~84, no.~4, p.~349, 2024.

\bibitem{Cognola:2007zu}
G.~Cognola, E.~Elizalde, S.~Nojiri, S.~D. Odintsov, L.~Sebastiani, and
  S.~Zerbini, ``{A Class of viable modified f(R) gravities describing inflation
  and the onset of accelerated expansion},'' {\em Phys. Rev. D}, vol.~77,
  p.~046009, 2008.

\bibitem{Golovnev:2022rui}
A.~Golovnev, ``{On the Role of Constraints and Degrees of Freedom in the
  Hamiltonian Formalism},'' {\em Universe}, vol.~9, no.~2, p.~101, 2023.

\bibitem{Blagojevic:2020dyq}
M.~Blagojevi\'c and J.~M. Nester, ``{Local symmetries and physical degrees of
  freedom in $f(T)$ gravity: a Dirac Hamiltonian constraint analysis},'' {\em
  Phys. Rev. D}, vol.~102, no.~6, p.~064025, 2020.

\bibitem{Bahamonde:2022kwg}
S.~Bahamonde, J.~Chevrier, and J.~Gigante~Valcarcel, ``{New black hole
  solutions with a dynamical traceless nonmetricity tensor in Metric-Affine
  Gravity},'' {\em JCAP}, vol.~02, p.~018, 2023.

\bibitem{McCrea:1992wa}
J.~D. McCrea, ``{Irreducible decompositions of non-metricity, torsion,
  curvature and Bianchi identities in metric affine space-times},'' {\em Class.
  Quant. Grav.}, vol.~9, pp.~553--568, 1992.

\bibitem{herrera}
R.~Gambini and L.~Herrera, ``{Einstein–Cartan theory in the spin coefficient
  formalism},'' {\em Journal of Mathematical Physics}, vol.~21, pp.~1449--1454,
  07 2008.

\bibitem{Golovnev:2023yla}
A.~Golovnev, ``{The geometrical meaning of the Weitzenb\"ock connection},''
  {\em Int. J. Geom. Meth. Mod. Phys.}, vol.~20, no.~Supp01, p.~2350219, 2023.

\bibitem{Aldrovandi:2013wha}
R.~Aldrovandi and J.~G. Pereira, {\em {Teleparallel Gravity}: {An
  Introduction}}.
\newblock Springer, 2013.

\bibitem{Blixt:2019ene}
D.~Blixt, M.~Hohmann, M.~Kr\v{s}\v{s}\'ak, and C.~Pfeifer, ``{Hamiltonian
  Analysis In New General Relativity},'' in {\em {15th Marcel Grossmann Meeting
  on Recent Developments in Theoretical and Experimental General Relativity,
  Astrophysics, and Relativistic Field Theories}}, 5 2019.

\bibitem{Golovnev:2023qll}
A.~Golovnev, ``{Geometry of teleparallel theories},'' 12 2023.

\bibitem{Bahamonde:2022zgj}
S.~Bahamonde and L.~J\"arv, ``{Coincident gauge for static spherical field
  configurations in symmetric teleparallel gravity},'' {\em Eur. Phys. J. C},
  vol.~82, no.~10, p.~963, 2022.

\bibitem{Zhao:2021zab}
D.~Zhao, ``{Covariant formulation of f(Q) theory},'' {\em Eur. Phys. J. C},
  vol.~82, no.~4, p.~303, 2022.

\bibitem{Hohmann:2021ast}
M.~Hohmann, ``{General covariant symmetric teleparallel cosmology},'' {\em
  Phys. Rev. D}, vol.~104, no.~12, p.~124077, 2021.

\bibitem{Corichi:2023ery}
A.~Corichi, J.~D. Reyes, and T.~Vukasinac, ``{On Covariant and Canonical
  Hamiltonian Formalisms for Gauge Theories},'' {\em Universe}, vol.~10, no.~2,
  p.~60, 2024.

\bibitem{Peldan:1993hi}
P.~Peldan, ``{Actions for gravity, with generalizations: A Review},'' {\em
  Class. Quant. Grav.}, vol.~11, pp.~1087--1132, 1994.

\bibitem{Hinterbichler:2012cn}
K.~Hinterbichler and R.~A. Rosen, ``{Interacting Spin-2 Fields},'' {\em JHEP},
  vol.~07, p.~047, 2012.

\bibitem{Krssak:2024xeh}
M.~Kr\v{s}\v{s}\'ak, ``{Teleparallel Gravity, Covariance and Their Geometrical
  Meaning},'' 1 2024.

\bibitem{Li:2011rn}
M.~Li, R.-X. Miao, and Y.-G. Miao, ``{Degrees of freedom of $f(T)$ gravity},''
  {\em JHEP}, vol.~07, p.~108, 2011.

\bibitem{Ferraro:2016wht}
R.~Ferraro and M.~J. Guzm\'an, ``{Hamiltonian formulation of teleparallel
  gravity},'' {\em Phys. Rev. D}, vol.~94, no.~10, p.~104045, 2016.

\bibitem{Guzman:2023oyl}
M.-J. Guzman, ``{The Hamiltonian constraint in the symmetric teleparallel
  equivalent of general relativity},'' 11 2023.

\bibitem{Heisenberg:2023tho}
L.~Heisenberg and M.~Hohmann, ``{Gauge-invariant cosmological perturbations in
  general teleparallel gravity},'' 11 2023.

\bibitem{Guzman:2020kgh}
M.-J. Guzman and S.~Khaled~Ibraheem, ``{Classification of primary constraints
  for new general relativity in the premetric approach},'' {\em Int. J. Geom.
  Meth. Mod. Phys.}, vol.~18, no.~supp01, p.~2140003, 2021.

\bibitem{Barker:2024grr}
W.~Barker and S.~Zell, ``{Einstein-Proca theory from the Einstein-Cartan
  formulation},'' {\em Phys. Rev. D}, vol.~109, no.~2, p.~024007, 2024.

\bibitem{Bajardi:2023vcc}
F.~Bajardi and S.~Capozziello, ``{Minisuperspace quantum cosmology in f(Q)
  gravity},'' {\em Eur. Phys. J. C}, vol.~83, no.~6, p.~531, 2023.

\bibitem{Miranda:2021oig}
M.~Miranda, D.~Vernieri, S.~Capozziello, and F.~S.~N. Lobo, ``{Effective
  actions for loop quantum cosmology in fourth-order gravity},'' {\em Eur.
  Phys. J. C}, vol.~81, no.~11, p.~975, 2021.

\bibitem{Bajardi:2020osh}
F.~Bajardi and S.~Capozziello, ``{$f(\mathcal {G})$ Noether cosmology},'' {\em
  Eur. Phys. J. C}, vol.~80, no.~8, p.~704, 2020.

\bibitem{Capozziello:2021pcg}
S.~Capozziello, A.~Finch, J.~L. Said, and A.~Magro, ``{The 3+1 formalism in
  teleparallel and symmetric teleparallel gravity},'' {\em Eur. Phys. J. C},
  vol.~81, no.~12, p.~1141, 2021.

\end{thebibliography}

\end{document}